\documentclass[aip]{revtex4-1}
\usepackage{graphicx}
\usepackage{amsmath}
\usepackage[version=4]{mhchem}
\usepackage{xcolor}
\usepackage[caption=false]{subfig}
\usepackage{multirow}

\newcommand{\ket}[1]{\left|#1\right\rangle}

\newcommand{\lacrev}{\textcolor{black}}
\draft 

\begin{document}

\title{Exploring Spin Symmetry-Breaking Effects for Static Field Ionization of Atoms: Is There an Analog to the Coulson-Fischer Point in Bond Dissociation?}

\author{Leonardo A. Cunha}
\email{leonardo.cunha@berkeley.edu}
\affiliation{Department of Chemistry, University of California, Berkeley, CA 94720 USA}
\affiliation{Chemical Sciences Division, Lawrence Berkeley National Laboratory, Berkeley CA 94720 USA}

\author{Joonho Lee}
\email{jl5653@columbia.edu}
\affiliation{Department of Chemistry, University of California, Berkeley, CA 94720 USA}
\affiliation{Department of Chemistry, Columbia University, New York, NY 10027 USA}

\author{Diptarka Hait}
\email{diptarka@berkeley.edu}
\affiliation{Department of Chemistry, University of California, Berkeley, CA 94720 USA}
\affiliation{Chemical Sciences Division, Lawrence Berkeley National Laboratory, Berkeley CA 94720 USA}

\author{C. William McCurdy}%
\email{cwmccurdy@lbl.gov}
\affiliation{Chemical Sciences Division, Lawrence Berkeley National Laboratory, Berkeley CA 94720 USA}
\affiliation{Department of Chemistry, University of California, Davis, CA 95616 USA}

\author{Martin Head-Gordon}
\email{mhg@cchem.berkeley.edu}
\affiliation{Department of Chemistry, University of California, Berkeley, CA 94720 USA}
\affiliation{Chemical Sciences Division, Lawrence Berkeley National Laboratory, Berkeley CA 94720 USA}

\date{\today}

\begin{abstract}
L\"owdin's symmetry dilemma is a ubiquitous issue in approximate quantum chemistry. In the context of Hartree-Fock (HF) theory, the use of Slater determinants with some imposed constraints to preserve symmetries of the exact problem may lead to physically unreasonable potential energy surfaces. On the other hand, lifting these constraints leads to the so-called \lacrev{broken symmetry} solutions that usually provide better energetics, at the cost of losing information about good quantum numbers that describe the state of the system. This behavior has been previously extensively studied in the context of bond dissociation. This paper studies the behavior of different classes of Hartree-Fock spin polarized solutions (restricted, unrestricted, generalized) in the context of ionization by strong static electric fields. We find that, for simple two-electron systems, UHF is able to provide a qualitatively good description of states involved during the ionization process (neutral, singly-ionized and doubly ionized states), whereas RHF fails to describe the singly ionized state. For more complex systems, even though UHF is able to capture some of the expected characteristics of the ionized states, it is constrained to a single $M_s$ (diabatic) manifold in the energy surface as a function of field intensity. In this case a better qualitative picture can be painted by GHF as it is able to explore different spin manifolds and follow the lowest solution due to lack of \lacrev{collinearity constraints} on the spin quantization axis.
\end{abstract}


\maketitle 

\section{\label{sec:intro}Introduction}

Discussions on the usefulness of symmetry-broken approximate solutions are a familiar topic in time-independent quantum chemistry.\cite{eisfeld2000detailed,sherrill1999performance,russ2004real,dobaczewski2000point,hait2019,hait2019beyond} The issue is encapsulated in what L\"owdin has called the ``symmetry dilemma"\cite{lowdin63}:  a more flexible trial wavefunction that does not necessarily preserve all of the symmetries of the exact Hamiltonian might lead to better energetics (\emph{i.e.} lower energy on account of the variational principle) at the cost of losing good quantum numbers that characterize the state of a given system. Within the single determinant Hartree-Fock (HF) model, classification of these symmetry-broken solutions is based on group theory considerations\cite{fukutome81}, but most commonly in electronic structure one uses terminology that reflects constraints imposed on the orbitals the comprise the single determinant\cite{paldus03}. For instance, requiring that both $\alpha$ and $\beta$ spin-orbitals share a common set of spatial functions (i.e. the electrons are paired whenever possible) leads to the well-known restricted closed-shell ($M_s = 0$) Hartree-Fock (RHF)  and the more general ($M_s \ne 0$) restricted open-shell HF (ROHF) models. RHF and ROHF are both eigenstates of total spin, $\hat{S}^2$, and its $z$ component, $\hat{S}_z$. Lifting this spin pairing constraint, such that $\alpha$ or $\beta$ spin orbitals can have different spatial functions, gives us the unrestricted Hartree-Fock (UHF) model, whose wavefunction is no longer an eigenfunction of $\hat{S}^2$. Further symmetry lowering by abolishing the notion of separate sets of $\alpha$ and $\beta$ orbitals leads to the generalized Hartree-Fock (GHF) wavefunction, which is not an eigenstate of either $\hat{S}^2$ or $\hat{S}_z$. 
Furthermore, number symmetry can also be relaxed following Hartree-Fock Bogoliubov theory,\cite{ebata2010canonical,bertsch2012symmetry} which yields a state that is not an eigenstate of the particle number operator ($\hat{N}$).

Within a finite basis, the HF energy, $E_{\text{HF}}(\boldsymbol{\theta})$, is a function of orbital rotation parameters, $\boldsymbol{\theta}$ that mix occupied and virtual orbitals.\cite{THOULESS1960225,szabo96,helgaker2014molecular} A solution of the Hartree-Fock equations zeros the orbital rotation gradient, such that $\nabla_{\boldsymbol{\theta}}E_\text{HF}=\mathbf{0}$ (i.e. it is guaranteed to be a stationary point). However a solution is not necessarily a minimum, and an analysis of the orbital Hessian, $E_{\text{HF}}^{\boldsymbol{\theta\theta}}$, is required to characterize the nature of a stationary point. If all of the eigenvalues of the Hessian are greater than zero, we have found a solution that is a local minimum, and is said to be stable. The question of which set of orbital parameters to include in the evaluation of the Hessian arises\cite{pople77}: if the HF solution is stable within the manifold determined by certain symmetry constraints (e.g. the RHF constraints), it is said that such a solution is internally stable. Lifting some or all of these symmetry constraints leads to characterization of the stationary point in manifolds of higher dimension than the one it was originally optimized in (e.g. characterizing an RHF solution in the space of UHF variations). This corresponds to an analysis of the external stability of the given solution. It might be useful to notice that stability analysis is closely connected with TDDFT/TDHF linear response equations\cite{mhg05,hait2019beyond} which are routinely available in standard quantum chemistry packages.

For common problems in ground-state electronic structure, internal stability analysis avoids convergence to spurious saddle points and excited states. On the other hand, external stability analysis sometimes reveals interesting physical insight into the nature of electron correlation. 
For instance, when single bonds are stretched beyond the so called Coulson-Fischer (CF) point\cite{coulson49}, the lowest triplet excited state (T$_1$) starts to mix with the singlet ground state (S$_0$) in a process known as the ``triplet instability",\cite{paldus67,toth2016finding} which leads to spin symmetry breaking. The resulting spin polarized UHF state has lower energy than the RHF state, but is no longer an eigenstate of $\hat{S}^2$: the state is said to be spin-contaminated due to its mixed singlet-triplet character. The RHF solution that pairs electrons in order to keep a well-defined $\hat{S}^2$ eigenstate fails to provide a qualitatively correct description of the dissociation process. The RHF pairing constraint, which fixes natural occupation numbers at 0 or 2 even at the dissociation limit (when one would expect the two orbitals to be singly occupied) leads to a state that spuriously preserves ionic character, which consequently fails to approach the correct asymptotic limit at complete dissociation. On the other hand, UHF is able to successfully provide a qualitatively accurate description of the ground state potential energy surface for single bond dissociation.\cite{szabo96} Similar considerations apply to molecules that are singlet diradicaloid\cite{stuck2011nature,jimenez2014polyradical,lee2019distinguishing} in character: because two orbitals have occupation numbers significantly different from two and zero, RHF cannot be qualitatively correct, while UHF exhibits spin contamination \lacrev{due to mixing of S$_0$ and T$_1$}. 

There are surely other situations in which similar symmetry dilemma arises, but where its consequences have not been so thoroughly explored as bond dissociation processes and singlet diradicaloids. One example lies in the interaction of atoms and molecules with strong electric fields.\cite{ivanov2005anatomy,reiss2008limits,tong2005empirical,holmegaard2010photoelectron,popruzhenko2014keldysh,shvetsov2016semiclassical,hartung2019magnetic,ludwig2014breakdown,litvinyuk2003alignment} Strong field chemistry and physics is a rapidly developing field, because of the rich array of new highly non-linear phenomena that emerge. For example, a strong oscillating field will nearly ionize bound electrons in one direction in a first half-cycle followed by reattachment and near ionization in the opposite direction in the second half-cycle. This leads to high harmonic generation (HHG),\cite{mairesse2010high,schafer1997high,christov1997high,itatani2005controlling,ghimire2019high} the emission of radiation by the driven bound system at frequencies which are many times that of the applied radiation. HHG and related phenomena are building blocks for the new field of attosecond science.\cite{krausz2009attosecond,ciappina2017attosecond,scrinzi2005attosecond,drescher2005attosecond} Therefore the use of quantum chemistry methods to model strong field phenomena is also attracting increasing interest.\cite{helgaker2015bfield,sato2020mp2,jagau2018cc,jagau2016barrier,schlegel2014tdci,schlegel2014tdcis2} However, to date there has been no systematic exploration of the role of symmetry-breaking in the mean field HF method in this context, to our knowledge. This work represents a first step in this direction, though the issue of inadequate field-free Hartree-Fock reference for real-time methods in strong field environments has been briefly mentioned in some previous works. \cite{kulander1987uhf,kvaal2012,kvaal2019,pedersen2020} 

By definition strong fields are those whose scale approaches the strength of internal electric fields experienced by the valence electrons of atoms and molecules (e.g. 1 a.u. for the H atom). Therefore the molecule and the field must be considered as a combined system, rather than treating the field as a perturbation. If such fields are static, then strong field ionization becomes possible, and some electrons may be unbound. Yet because the molecule-field system is treated as a whole, the symmetry dilemma may arise for at least some values of the applied field. In this sense, the magnitude of the field is a control variable similar to the degree of bond-stretching in the dissociation of a closed-shell stable molecule. This paper explores the role of symmetry-breaking in HF solutions for atom-field systems where a static field is scanned across values that can strip one or more electrons from the atom. This investigation is interesting in its own right, and also may help to set the stage for a subsequent analysis in the context of time-dependent fields. Our paper is arranged as follows: in Section \ref{sec:model} we discuss our approach to approximate the description of continuum-like states using ghost basis functions, an analysis of the HF spin symmetry-broken solutions for the static field ionization of helium and neon is presented in Sections \ref{sec:he} and \ref{sec:neon}, respectively, and in Section \ref{sec:conclusions} we assess our main conclusions while presenting an outlook of future work.

\section{\label{sec:model}Static Field Ionization in a Finite Basis}
\begin{figure}[!htp]
    \centering
    \includegraphics[scale=0.25]{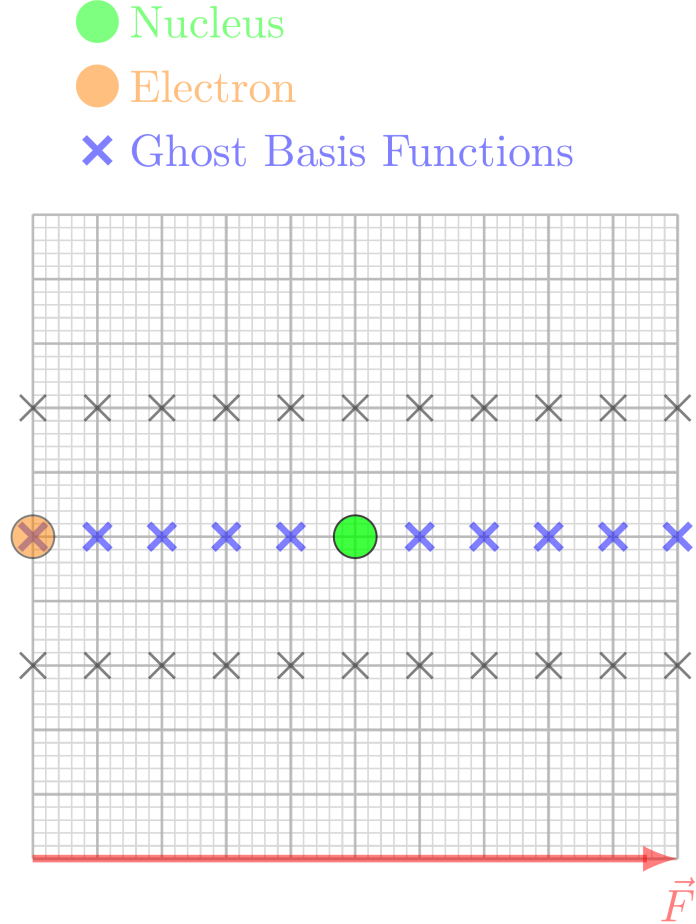}
    \caption{Setup of our model for static field ionization. A series of ghost functions was added passing through the atom and in the direction of the field, allowing electron density to escape/ionize from the system (the grey crosses represent additional race track ghost functions, as discussed in Sec.~\ref{sec:neon}).}
    \label{fig:model_setup}
\end{figure}

The first issue that arises when applying an electric field to an atomic or molecular system is that, strictly speaking, bound states are no longer supported by the combined potential arising from the system and the field.\cite{nimrod2020,brown2015properties} These states are now resonances and a suitable discretization for both \lacrev{localized} and continuum states are needed to describe them. From a theoretical perspective, the scattering community has developed a wide range of tools to treat these resonance states, such as exterior complex scaling (ECS) \cite{simon1979definition, scrinzi1993finite,rescigno1997making,he2007absorbing,rom1990tunneling,morgan1981calculation,McCurdy2004ecs,ShihI2013ecs}, complex absorbing potentials (CAP) \cite{masahiro2015caps,zuev2014complex,lefebvre2005resonance, santra2006complex,vibok1992parametrization,jagau2014fresh,santra2002complex,feuerbacher2003complex,muga2004complex}, and different grids as schemes to discretize space \cite{manthe1996time,Sanchez1997bspline, Martin1999bspline,Bachau2001spline,littlejohn2002general,tao2009grid,McCurdy2020dvr}. On the other hand, the quantum chemistry community has long advocated for the use of atomic orbital (AO) expansions based on Gaussian functions as an efficient way to compute ground and excited state properties of molecular systems. Since our goal is to study static field ionization exploring the common toolbox of quantum chemistry, we shall resort to the usual finite AO basis set treatment. What are the consequences of such choice? The most concerning one is that ionization losses are hindered, since electronic density will be constrained near the atom/molecule. Two possible ways\cite{Luppi:2012} of circumventing this issue are the use of highly diffuse basis sets and discretizing the relevant part of the space by adding a ghost atoms/functions that could be populated by electron density as ionization takes place. We have opted for this second option and added a series of ghost functions in a line passing through the atom and in the direction of the applied field. This allows us to capture some of the effects of ionization as electron density can escape the system and populate the discretized space (Fig.~\ref{fig:model_setup}). \lacrev{This setup is inspired by previous works that aimed to study the real-time electron dynamics of systems in the context of strong fields and high-harmonic generation (HHG), where the addition of diffuse and ghost functions to the basis set is important for a good description of the Rydberg and unbound states of the system, respectively.\cite{Luppi:2012}}

Given this setup for our problem, we also need to interpret what the exact results would look like in this finite basis set approach. We start our analysis by noticing that, assuming an adequate discretization of the space by the ghost centers, we can separate the problem/potential into two parts: one associated with the molecular/atomic system and another related to the artificial box that is effectively created by the extent of the ghost centers. Introductory quantum mechanics teaches us that, to first order, the energy levels of a charged particle in a box subjected to a uniform static electric field decrease linearly with the strength of the field. \cite{landauqm} Moreover, increasing the field strength affects the molecular potential by suppressing the barrier to ionization for higher excited state (especially Rydberg states) and increasing the tunneling probability for low-lying states. In this sense, higher excited states easily become continuum-like states and ``dive" down in energy as the field becomes stronger. At certain field strengths, these continuum-like states start to interact with the low-lying bound states and an avoided crossing is observed: the former bound state acquires continuum-like character by localizing around the edge of the artificial box, with an energy that decreases linearly as the field strength continues to increase. Notice that the applied electric field turns all of the states of the system into resonances with finite lifetime.\cite{landauqm} and a proper discussion on how to obtain DC Stark lifetimes for several molecular system using quantum chemistry methods has already been presented elsewhere.\cite{jagau2016barrier,jagau2018cc,jagau2020}.

\begin{figure}[!htp]
    \centering
    
    \subfloat[Singlet states\label{fig:he_asci_s}]{%
         \includegraphics[scale=0.75]{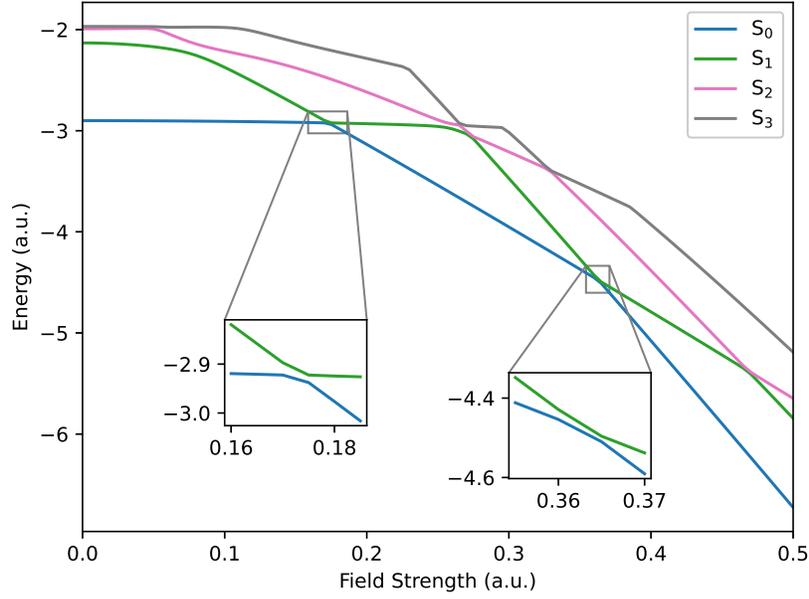}}
    
    \subfloat[Triplet states\label{fig:he_asci_t}]{%
         \includegraphics[scale=0.75]{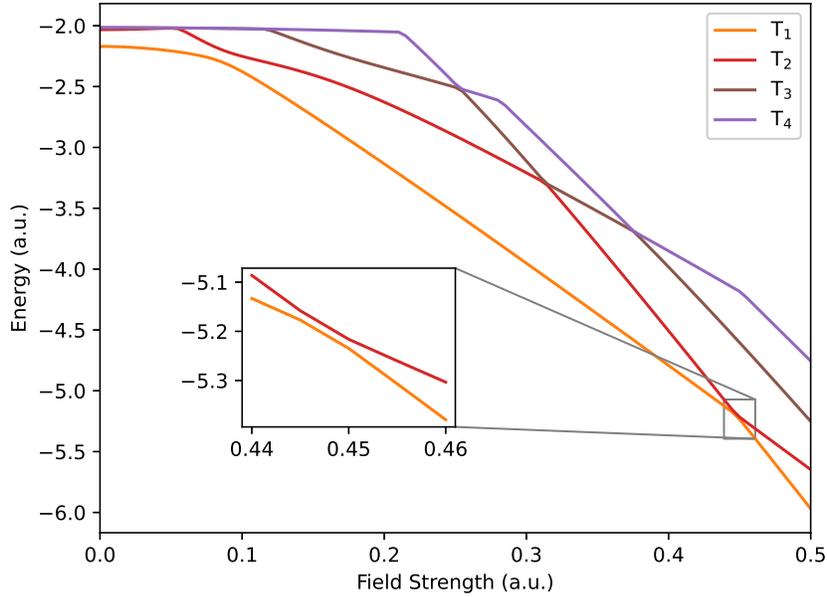}}   
    \caption{FCI (exact) results for the static field ionization of He for (a) lowest four singlet states, (b) lowest four triplet states. The ``kinks" correspond to avoided crossings between continuum-like states associated to the artificial finite box and localized states from the atom. For the ground state, the crossings also give us the signature associated to ionization events. The atom is described by the aug-cc-pVTZ basis set, whereas each ghost center is described by hydrogenic STO-3G 1s function separated by 0.5 $\textrm{\AA}$.}
    \label{fig:he_exact}
\end{figure}

Fig.~\ref{fig:he_exact} illustrates this process for a He atom (described by the functions contained in the aug-cc-pVTZ basis set\cite{dunning1989gaussian,dunning1992gaussian}) in an artificial box comprised of 18 hydrogen STO-3G\cite{stewart1970sto} s-type ghost centers spaced by 0.5 \textrm{\AA} spanning a range of 4.5 $\textrm{\AA}$ around the atom \lacrev{(which should be adequate for an unambiguous determination of the state of ionization of the system given that this distance is much larger than the atomic radius of He). We shall focus on the behavior of the ground-state S$_0$ and the first singlet excited state S$_1$ as a function of field strength.} Initially, in the absence of the field, \lacrev{these states are well separated in energy and there is no mixing between them}. As the field strength increases, we observe the characteristic quadratic polarization effect on the energy (which can be described by perturbation theory).\cite{hait18} Higher lying states are more polarizable, so this effect becomes more evident. These high energy excited states are also more easily ionized and the transition between the quadratic behavior characteristic of a bound state to a linear dependence of the energy on field strength happens at lower fields. \lacrev{ At F = 0.175 a.u., we observe an avoided crossing between the first singlet state (which has been previously ionized) and the unionized ground state. This leads to a change in character of the ground state after the avoided crossing: initially unionized, with both electrons localized around the atom, the ground state becomes singly ionized, with one electron detaching from the atom and localizing at the edge of the box. The second ionization is characterized by yet another avoided crossing between the singly ionized S$_0$ and doubly ionized S$_1$ at a field of F = 0.36 a.u.} Between these two field strengths, we note that the singly ionized T$_1$ state is degenerate with the ionized ground state S$_0$. It is worth noticing that the potential energy curves presented here as a function of field strength are very similar to stabilization graphs proposed by Simons\cite{simons1981stab} and widely used to describe temporary ions and other atomic/molecular resonances\cite{jordan2021res}. There, information about the resonance's lifetime can be obtained by an analysis of the avoided crossing \lacrev{that arise by changing the variational parameters associated with the basis set. Here, however, the basis set is fixed and the energy is plotted as a function of varying field strength.}

Our goal is to understand how different approximations, such as HF and MP2 can recover this exact finite basis behavior. We shall put emphasis on the effects of allowing for spin polarization (\emph{i.e.}, difference between using restricted and unrestricted orbitals) to describe field ionization as an analog to what happens in bond dissociation of molecules. All calculations were performed with the Q-Chem 5.3 package\cite{QCHEM4} following the same protocol: we used the aug-cc-pVTZ basis set for the atomic center and hydrogenic STO-3G s functions for the ghost centers. We also performed internal stability analysis in order to ensure that the solution within the domain of each HF class (RHF, UHF and GHF) was the lowest possible.

\section{\label{sec:he}Field Ionization of He}

\subsection{\label{sec:minba_he}Hartree-Fock (HF) Minimum Basis Model}
We start our discussion about the relation between spin polarization and static field ionization by analyzing the behavior of the Hartree-Fock (HF) solution in a simple, yet instructive, minimum basis model of a two electron system. This model is comprised of a single basis function localized around the atom (represented by $\ket{A}$ in Table~\ref{tab:orb_dets_ru}) and a single function (represented as $\ket{C}$ in Table~\ref{tab:orb_dets_ru}) placed at a distance $d$ from the atom in the direction of the applied field to represent the discretized continuum and that can support the flux of ionized electrons once the field is switched on. It should be noted that this is analogous to the toy model used to describe the bond stretching of H$_2$ in the minimum basis \cite{hait2019beyond,szabo96}. Here, we have also assumed that the distance $d$ is large enough such that there is no overlap between the orbitals representing the atom and the discretized continuum. Moreover, we assumed that the ghost function is diffuse enough that we can neglect its kinetic energy ($\langle C | \hat{T} | C \rangle \approx 0$), but not too diffuse relative to the distance $d$. We can then construct the HF orbitals as linear combinations of these two functions.

In this toy model, the RHF solution is controlled by a single variational parameter $\theta$ that mixes the atomic basis function with the continuum-like ghost function (Table~\ref{tab:orb_dets_ru}). Analyzing the stationary and stability conditions of the energy obtained as the expectation value of the Hamiltonian with the RHF determinant leads to two limiting cases. The solution characterized  by the double occupancy of the atom-centered orbital ($\theta = 0 $) is found to be stable when the product of the applied field strength ($F$) and the distance between the basis functions ($d$) is less than the $IP_1$, the first ionization potential of the system (\emph{i.e.} $Fd < IP_1$). On the other hand, the $\theta = \frac{\pi}{2}$ solution, which corresponds to 2 electrons paired up on the ghost function that supports the ionized flux, is stable when $Fd > IP_2$, where $IP_2$ is the second ionization potential of the system. At intermediate field strengths, $IP_1 < Fd < IP_2$, the stable solution is given by a state that has mixed neutral and doubly ionized character. It is interesting to notice that RHF only allows for the combined ionization of the electron pair and the intermediate state associated with single ionization is never reached.  Hence, we see that restricted HF (RHF) cannot recover the exact behavior shown in Fig.~\ref{fig:he_exact} for the ground state, as it only connects the neutral solution (with the electrons paired up on the atom) and the doubly ionized state.

\begin{table}[!htb]
    \centering
    \begin{tabular}{ |c | c | c|}
        \hline
         \multicolumn{3}{|c|}{Model Assumptions}\\ \hline
         \multicolumn{3}{|l|}{1a. $\ket{A}$ - atomic spatial orbital} \\
         \multicolumn{3}{|l|}{1b. $\ket{C}$ - continuum-like orbital} \\
         \multicolumn{3}{|l|}{2. $\langle A | C \rangle \approx 0$} \\
         \multicolumn{3}{|l|}{3. $\langle C | \hat{T} | C \rangle \approx 0$} \\ \hline
         \multicolumn{3}{|c|}{Spatial Orbitals}\\ \hline
         \multicolumn{3}{|c|}{$\ket{1} =  \cos\theta_1 \ket{A} + \sin\theta_1 \ket{C}$} \\
         \multicolumn{3}{|c|}{$\ket{2} =  \cos\theta_2 \ket{A} + \sin\theta_2 \ket{C}$} \\ \hline
         \multicolumn{3}{|c|}{Stable RHF - $\ket{\Psi} = \ket{1\bar{1}}$}\\ 
         \hline
         $\theta = 0$ & $0 < \theta < \frac{\pi}{2}$ & $ \theta = \frac{\pi}{2}$ \\
         $Fd < IP_1$ & $IP_1 < Fd < IP_2$ & $Fd > IP_2$ \\ 
         $\ket{\Psi} = \ket{A\bar{A}}$ & mixed solutions & $\ket{\Psi} = \ket{C\bar{C}}$ \\ \hline
         \multicolumn{3}{|c|}{Stable UHF - $\ket{\Psi(\theta_1,\theta_2)} = \ket{1\bar{2}}$}\\ \hline
         $\theta_1 = \theta_2 = 0$ & $\theta_1 = 0, \theta_2 = \frac{\pi}{2}$ or $\theta_1 = \frac{\pi}{2}, \theta_2 = 0$ & $\theta_1 = \theta_2 =  \frac{\pi}{2}$ \\
         $Fd < IP_1$ & $IP_1 < Fd < IP_2$ & $Fd > IP_2$ \\ 
         $\ket{\Psi} = \ket{A\bar{A}}$ & $\ket{\Psi} = \ket{C\bar{A}}$ or $\ket{\Psi} = \ket{A\bar{C}}$ & $\ket{\Psi} = \ket{C\bar{C}}$ \\ \hline
    \end{tabular}
    \caption{Analytic representation of molecular orbitals and ionization states predicted by HF for minimal basis model. $\ket{{A}}$ and $\ket{{C}}$ are the atomic and ghost/continuum basis functions, respectively. The stationary and stability conditions for RHF and UHF lead to different character of the solutions as a function of field strength and UHF is the only model capable of describing single ionization.}
    \label{tab:orb_dets_ru}
\end{table}

The analysis of the unrestricted case is more intricate and interesting. Two independent variational parameters ($\theta_1,\theta_2$) control the mixing of the basis functions for the independent sets of $\alpha$ and $\beta$ orbitals (Table~\ref{tab:orb_dets_ru}). The stationary condition on the energy expectation value for the UHF determinant leads to four different solutions for the $(\theta_1,\theta_2)$ pairs. The solutions $(\theta_1,\theta_2) = (0,0)$ and $(\theta_1,\theta_2) = (\frac{\pi}{2},\frac{\pi}{2})$ reduce to RHF, corresponding to double occupation of the atomic and ghost orbitals, respectively. The former case is stable when $Fd < IP_1$, whereas the stability for the latter is achieved when $Fd > IP_2$. The difference in the UHF case is the possibility of achieving intermediate stable solutions $(\theta_1,\theta_2) = (0,\frac{\pi}{2})$ and $(\theta_1,\theta_2) = (\frac{\pi}{2},0)$ for the intermediate range of field strengths  $IP_1 < Fd < IP_2$. These extra UHF solutions are characterized by having one electron occupying the atomic orbital and the other one localized on the ghost function. Therefore, we see that UHF is able to characterize the singly ionized state. This ability, however, comes at a cost: in an analogy to the common problem of bond dissociation at the UHF level, these single determinant unrestricted solutions for the intermediate range of field strengths are composed of an equal mixing of singlet and triplet configurations that results in a spin-contaminated state with $\langle S^2 \rangle = 1$\cite{szabo96}. Nonetheless, UHF is qualitatively able to describe the static field ionization behavior expected from our analysis of the exact results (Fig.~\ref{fig:he_exact}), showing the three distinct regimes corresponding to the three possible levels of ionization. In addition, the large separation between $\ket{A}$ and $\ket{C}$ leads to a vanishing singlet-triplet gap, so the contamination has no direct impact on the energetics. \lacrev{Finally, we note that the 1:1 quantitative mapping between the position of the ``kinks" in the energy curve as a function of field strength and ionization potentials (IPs) is only possible if the conditions outlined in Table~\ref{tab:orb_dets_ru} are satisfied. The model can be easily modified to account for the electrostatic repulsion between $\ket{A}$ and $\ket{C}$ and the kinetic energy of the populated ghost function ($\langle C | \hat{T} | C \rangle$, leading to a better numerical agreement for the IPs. Alternatively, a decomposition of the total energy into atomic, continuum and interaction contributions could lead to better quantitative agreement for the IPs. Qualitatively, however, the main features do not depend on the specific nature of the ghost functions.}

\subsection{\label{sec:bas_he}Wavefunction Methods in a Larger Basis}
\begin{figure}[!htb]
    \centering
    \subfloat[SCF total energy,\label{fig:he_hf_energy}]{%
         \includegraphics[scale=0.52]{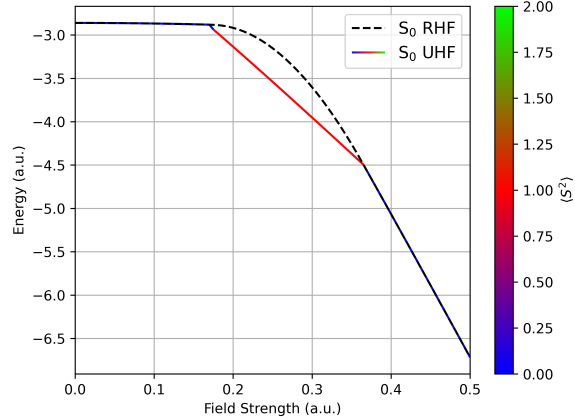}
    }

    \subfloat[Dipole moment,\label{fig:he_hf_dipole}]{%
         \includegraphics[scale=0.52]{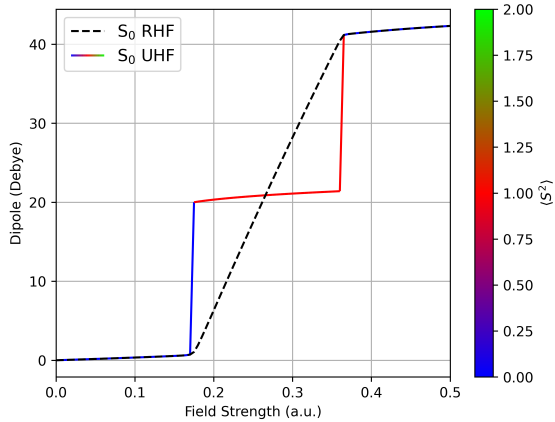}
    }
    \hfill
    \subfloat[Mulliken charges on He,\label{fig:he_hf_charge}]{%
         \includegraphics[scale=0.52]{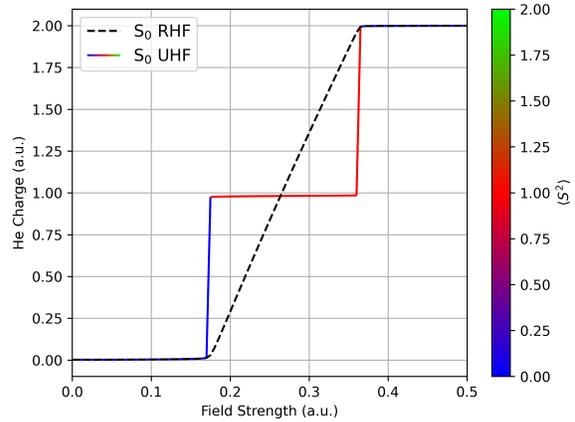}
    }
    \caption{ (a) SCF total energy, (b) Dipole moment and (c) Mulliken charges on He as a function of field strength for the RHF and UHF solutions. In all cases, the RHF curve smoothly changes as the field strength increases, while UHF presents the characteristic kinks that are associated to ionization events.}
    \label{fig:he_all}
\end{figure}

We move on to an analysis of the performance of other wavefunction based methods on a larger basis set to describe the static field ionization of He. We should point out that our emphasis is, once again, on the difference between how different flavors of self-consistent field (SCF) solutions can capture ionization effects in this discretized continnum model. In this case, He is represented by the larger aug-cc-pVTZ basis and the continuum is represented by a track of \lacrev{18} 1s hydrogenic STO-3G ghost basis functions spanning the range between -4.5 \AA~ and 4.5 \AA~ (with equal spacing of 0.5 \AA~ between each ghost center) along the direction of the field. Fig.~\ref{fig:he_hf_energy} summarizes the results in this larger basis set.

Just as expected from the analytical model, RHF cannot qualitatively describe the correct behavior for single-electron ionization due to the constraint that the two electrons are paired. For field strengths smaller than 0.175 a.u., we observe the neutral (two electrons localized around the atom) ground state, and for field strengths greater than 0.36 a.u., we observe the doubly ionized state of the atom. We notice a smooth transition between neutral atom and its atom dication state for intermediate field strengths. This can also be observed when we analyze how the total electric dipole (Fig.~\ref{fig:he_hf_dipole}) of the system and the charge of the He atom (Fig.~\ref{fig:he_hf_charge}) change as a function of the applied field. UHF, on the other hand, captures the essential features expected from the exact solution: the energy curve shows three distinct regions separated by first derivative discontinuities. These kinks are associated with spin-polarization effects as illustrated by the transition between $\langle S^2 \rangle = 0$ to $\langle S^2 \rangle = 1$. The sharpness of this transition (\emph{i.e.}, how fast the spin polarization occurs) is highly dependent on the size of the basis set: for smaller basis sets, $\langle S^2 \rangle$ seems to continuously change from 0 to 1, which indicates another remarkable similarity between the analysis presented in this work and the nature of the Coulson-Fischer point for bond dissociation. For larger basis sets, the spin polarization transitions appears to be much more sudden and discontinuous. We next consider the behavior of electric dipole and He charge for the UHF solution (Figs.~\ref{fig:he_hf_dipole} and~\ref{fig:he_hf_charge}). \lacrev{We do however note that the numerical magnitude of the change in the dipole moment is directly proportional to the size of the box created by the ghost functions.} Both observables indicate that UHF qualitatively captures the three distinct regimes expected for the ionization of helium: the polarization of the electron pair on the neutral atom, the singly ionized state and the doubly ionized state. Again, the ability of UHF to describe this singly ionized state comes at the cost of spin polarization as indicated by $\langle S^2 \rangle= 1$ for field strengths in the range that generates one unit of positive charge on helium.

\begin{figure}
    \centering
    \includegraphics[scale=0.65]{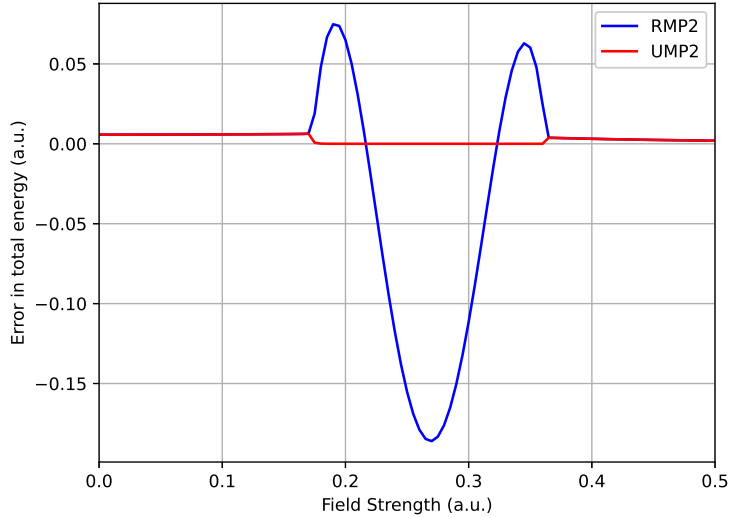}
    \caption{RMP2 and UMP2 results for the static field ionization of He. Due to the poor performance of the RHF reference, RMP2 presents energies lower than the exact ones, indicating that the MP2 amplitudes are overcompensating for the bad HF reference in the singly ionized regime. On the other hand, UMP2 is able to correctly recover almost exactly the FCI.}
    \label{fig:he_mp2}
\end{figure}

Moreover, we briefly analyze the MP2 (Fig.~\ref{fig:he_mp2}) performance for strong field ionziation of He using restricted (RMP2) and unrestricted (UMP2) orbitals. As expected, since UHF is a better reference to describe all possible charge regimes for our systems, UMP2 recovers all of the qualitative features of the exact solution. By contrast, RMP2 seems to oscillate above and below the exact solution: for field strengths closer to the ionization points, RMP2 energies are, as expected, above the exact values and seem, at first, to follow the same pattern as the RHF curve. However, RMP2 drops below the exact curve for fields in the mid-range for the singly ionized state. This indicates the RMP2 is over-correcting for the bad reference provided by the restricted orbitals in this regime. Nonetheless, this behavior is not as catastrophic as in the RMP2 potential energy curve for the dissociation of H$_2$, where incipient orbital degeneracies cause a divergent PES as the bond length increases.\cite{szabo96}   Although not explored in the present work, this failure could be a fertile ground for an investigation of the efficiency of orbital-optimized MP2 schemes (OO-MP2) with regularization\cite{lee2018oomp2} to cheaply but accurately recapture some of the correlation effects associated to field ionization.

Our final analysis for the He atom is based on an attempt to remove most of the electric field and spurious discretized continuum contributions to the total energy, accounting only for the purely electronic energy of the atom as a function of the field strength. Within the dipole approximation, we defined an internal energy operator $\hat{U}$:
\begin{equation}\label{eqn:int_energy}
    \hat{U} = \hat{H}_{elec} - \hat{\vec{\mu}} \cdot \vec{F}
\end{equation}
The electronic energy of the combined atom-continuum system can then be calculated via:
\begin{equation}\label{eqn:elec_energy}
        E_{elec} = \langle {\hat{H}_{elec}} \rangle = E^{atom}_{elec} + E_{elst} + E^{ghost}_{elec} = \langle \hat{U} \rangle + \langle \hat{\vec{\mu}} \rangle \cdot \vec{F}
\end{equation}
Eq.~\ref{eqn:elec_energy} includes three main components: (i) the electronic energy of the isolated atom ($E^{atom}_{elec}$), (ii) the electrostatic interaction between discretized continuum and He atom ($E_{elst}$), and (iii) the electronic energy of the populated discretized continuum ($E^{ghost}_{elec}$).

In the limit of two non-interacting subsystems (the isolated atom and the discretized continuum represented by the ghost functions), the super-system wavefunction is given by the product of the two individual wavefunctions and the super-system density matrix ($D$) can be written as a direct sum of the atomic density matrix ($D^A$) and a density matrix associated with the ghost functions ($D^G$). For our setup, however, the definition of the atomic subsystem is somewhat fuzzy since, in the complete basis set limit for the atom, the most diffuse basis functions will overlap with the nearest ghost centers added to represent the continuum electrons. This situation could be ameliorated by increasing the distance between this nearest ghost center and the atom or by noticing that, after ionization, electrons will tend to occupy the ghost functions closer to the edge of the artificial box used for the discretization of the continuum states. This allows us to have a clear definition of the ghost subsystem and we can then define the atomic contribution as the difference between the total energy and the energy contributions from the ghost basis functions.

By following this approach, $D^A$ and $D^G$ are defined as blocks of the super-system density matrix associated with the electron density represented by basis functions localized around the atom and  with the density represented by basis functions centered in the track of ghost centers, respectively. The electrostatic interaction can then be calculated as 
\begin{equation}\label{eqn:elst_energy}
E_{elst} = \sum_{\mu,\nu \in G} D^G_{\mu\nu}V^{G}_{\mu\nu} + \sum_{\mu,\nu \in G}\sum_{\lambda,\sigma \in A}D^G_{\mu\nu}(\mu\nu|\lambda\sigma)D^A_{\sigma\lambda} 
\end{equation}
In Eq~\ref{eqn:elst_energy}, we separated the contribution of the attractive interaction between the electrons populating a ghost function and the He nucleus (which is accounted by the ghost function block of the nuclear attraction one-electron integrals, $V^G$) and the repulsive screening between the continuum-like electron density ($D^G$) and the remaining electron density localized around He ($D^A$). 

The electronic energy of the populated discretized continuum is expressed as a trace over its subsystem density matrix, $D^G$:
\begin{equation}\label{eqn:embd_ghost_energy}
    E^{ghost}_{elec} = \frac{1}{2}\sum_{\mu,\nu \in G} (F^G_{\mu \nu} + T^G_{\mu \nu}) D^G_{\mu\nu}
\end{equation}
where $T^G$ is the kinetic component of the one-electron integrals for the ghost basis functions, F$^G$ is a Fock matrix in which we replaced the usual one-electron integrals by only its kinetic component (since for the discretized continuum subsystem there is no nuclear attraction). 

Eq. \ref{eqn:embd_ghost_energy} is clearly correct in the non-overlapping limit. In the presence of overlap, Eq. \ref{eqn:embd_ghost_energy} corresponds to the embedded mean-field theory\cite{fornace2015embedded, ding2017embedded}
decomposition of the total energy of two composite systems (in our case He and ghost centers) into the energies of each subsystem and their interaction energy. \lacrev{It is worth noticing that we have not explicitly included other energy contributions arising from the mixed atom-ghost density contributions or atom-ghost exchange effects in Eq.~\ref{eqn:elec_energy}. Implicitly, these contributions are accounted for in the definition of $E^{atom}_{elec}$ (Eq.~\ref{eqn:atom_energy}) and are small due to large separation between the atom and the ghost function on the edge of the track.}

Finally, the quantity of interest $\Delta E_{elec}^{atom}$ is given in terms of the atomic subsystem electronic energy at a given field strength
\begin{equation}\label{eqn:atom_energy}
E^{atom}_{elec}(F) = \langle \hat{U} \rangle + \langle \hat{\vec{\mu}} \rangle \cdot \vec{F} - E_{elst} - E^{ghost}_{elec}
\end{equation}
relative to the corresponding quantity at zero field: 
\begin{equation}\label{eqn:deltaE_ip}
    \Delta E^{atom}_{elec} = E^{atom}_{elec}(F) - E^{atom}_{elec}(F=0)  
\end{equation}

\begin{figure}
    \centering
    \includegraphics[scale=0.75]{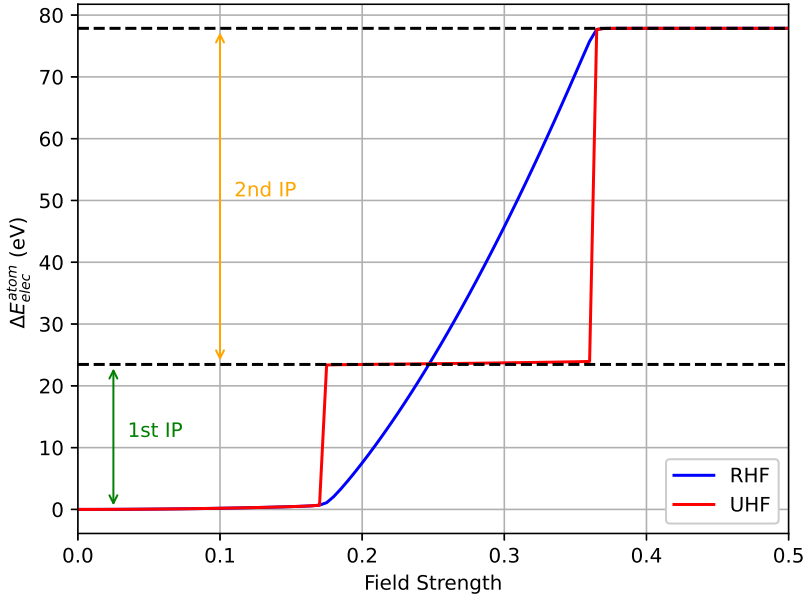}
    \caption{$\Delta E^{atom}_{elec}$ as defined in Eq~\ref{eqn:deltaE_ip}. UHF is able to correctly recover information about both first and second ionization potentials of He, whereas RHF, missing a proper description of the first ionization event, only captures the second IP. \lacrev{The IP's were calculated at the HF/aug-cc-pVTZ+ghost track level of theory.}}
    \label{fig:he_dEip}
\end{figure}

\begin{figure}
    \centering
    \centering
    \subfloat[NOONs for He + ghost system \label{fig:he_uhf_no}]{%
         \includegraphics[scale=0.52]{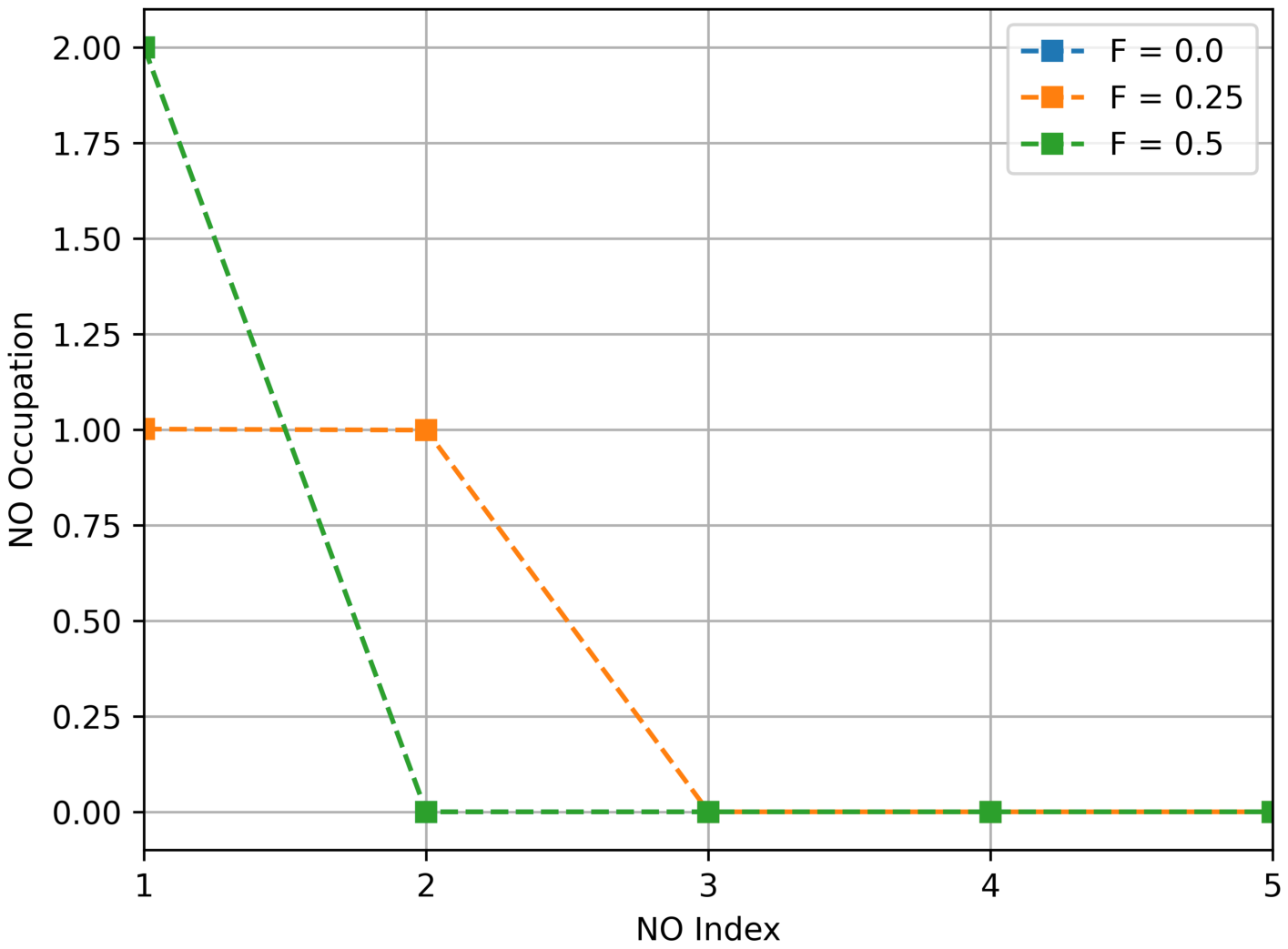}
    }
    \hfill
    \subfloat[NOONs for He subsystem \label{fig:he_uhf_nos}]{%
         \includegraphics[scale=0.52]{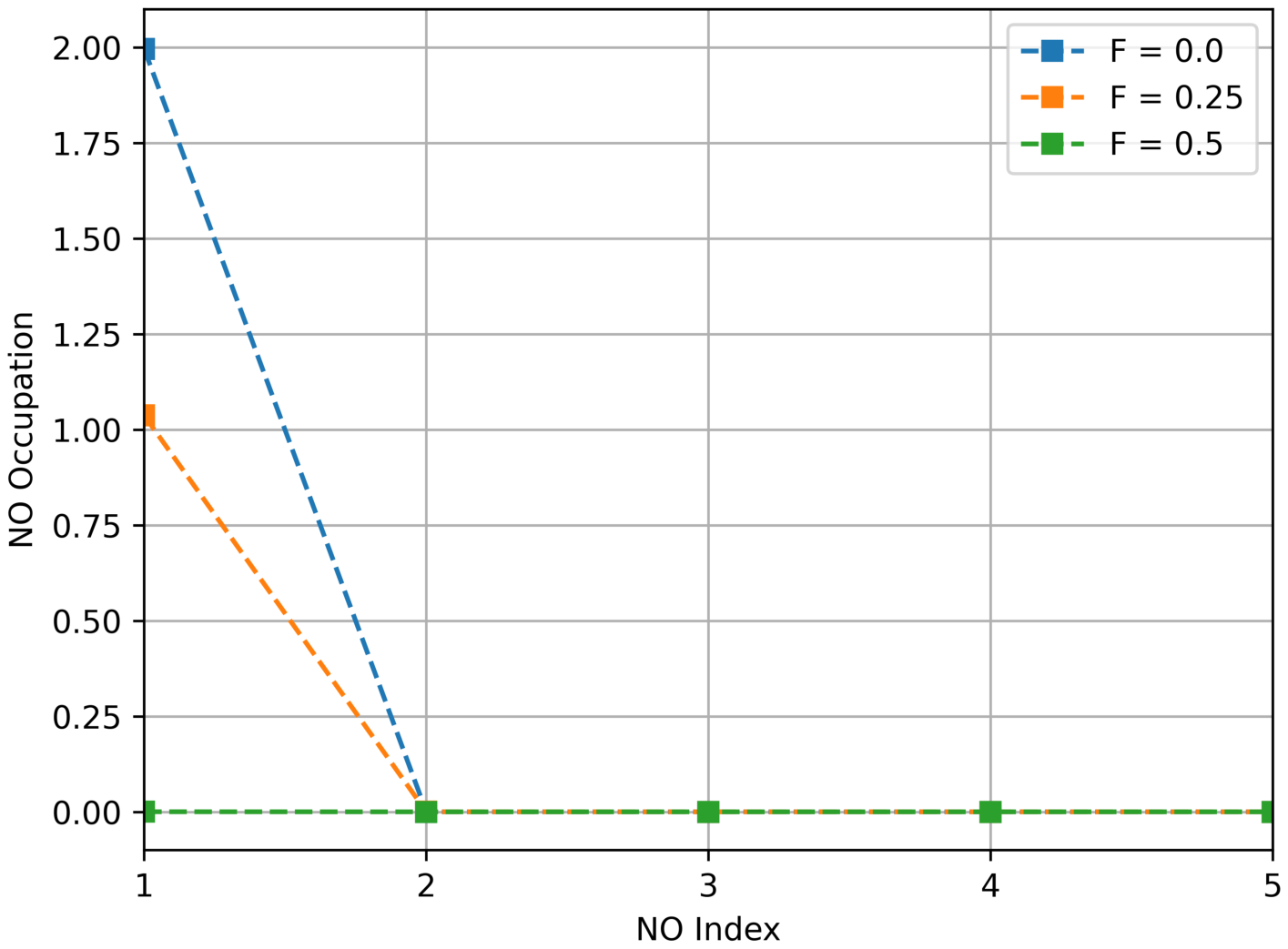}
    }
    
    \subfloat[NOONs for ghost subsystem \label{fig:he_uhf_nog}]{%
         \includegraphics[scale=0.52]{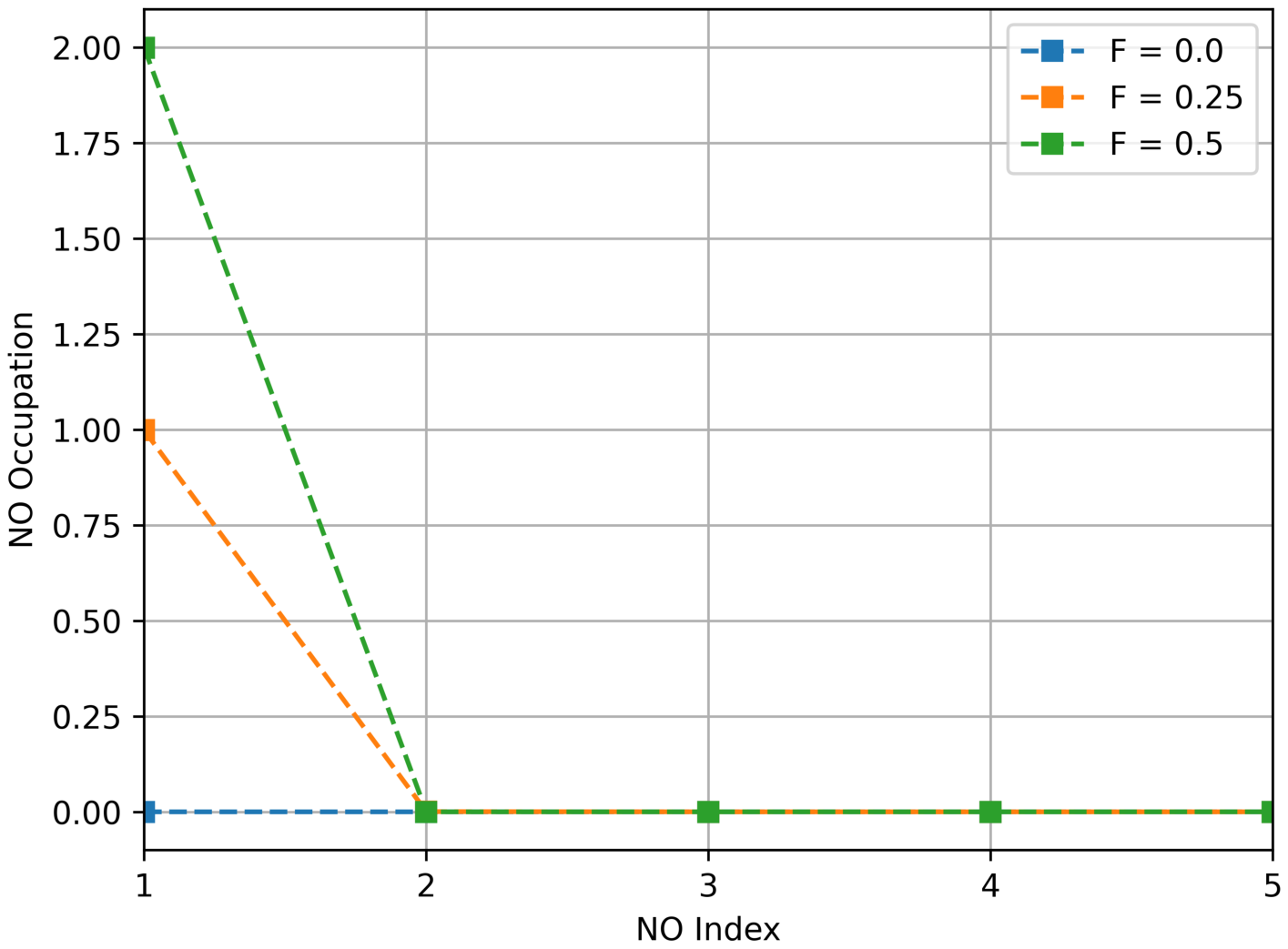}
    }
    \hfill
     \subfloat[Spin density for F = 0.25 a.u. \label{fig:he_uhf_spind}]{%
         \includegraphics[scale=0.52]{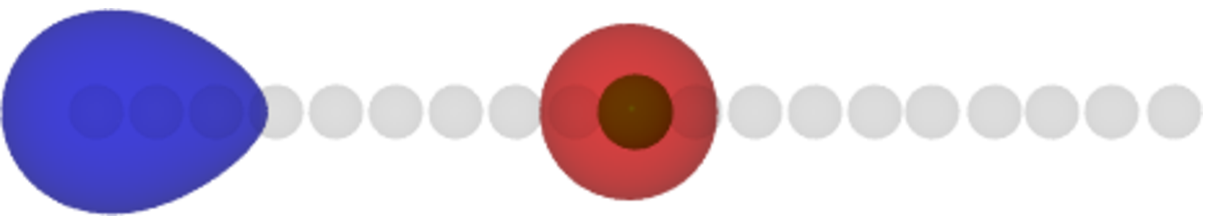}
    }
    
    \caption{Natural Orbital Occupation Numbers (NOONs) for (a) He + Ghost complex system, (b) He subsystem (c) Ghost subsystem for the ionization of He in different regimes: neutral (F = 0 a.u.), singly ionized (F = 0.25 a.u.) and doubly ionized (F = 0.5 a.u.). (d) Spin density for the singly ionized state of He.}
    \label{fig:he_no_all}
\end{figure}

Based on our previous analysis, this approach should be more suitable for the UHF solution in which the separable density assumption holds to a greater extent and as it can properly account for ionization. Fig.~\ref{fig:he_dEip} shows that the kinks in the total UHF energy (Fig.~\ref{fig:he_hf_energy} and the dipole and charge discontinuities for the UHF solution (Figs.~\ref{fig:he_hf_dipole} and~\ref{fig:he_hf_charge}) are, once again, intrinsically related to ionization events, as we can directly recover information about the first and second ionization potentials (represented by the dashed lines in Fig.~\ref{fig:he_dEip}) of He by removing the spurious field and ghost functions contributions to the total energy. \lacrev{Finally, the analysis based on the decomposition of the density matrix into its atomic and ghost contributions allows us to obtain more information about the electronic structure of each subsystem: an eigenvalue decomposition of the combined atom-ghost, atomic and ghost density matrices indicates the number of electrons for each subsystem as illustrated in Fig.~\ref{fig:he_no_all}.}

\section{\label{sec:neon}Field Ionization of a neon atom}
\begin{figure}[!htb]
    \centering
    \subfloat[SCF total energy \label{fig:ne_hf_energy}]{%
         \includegraphics[scale=0.52]{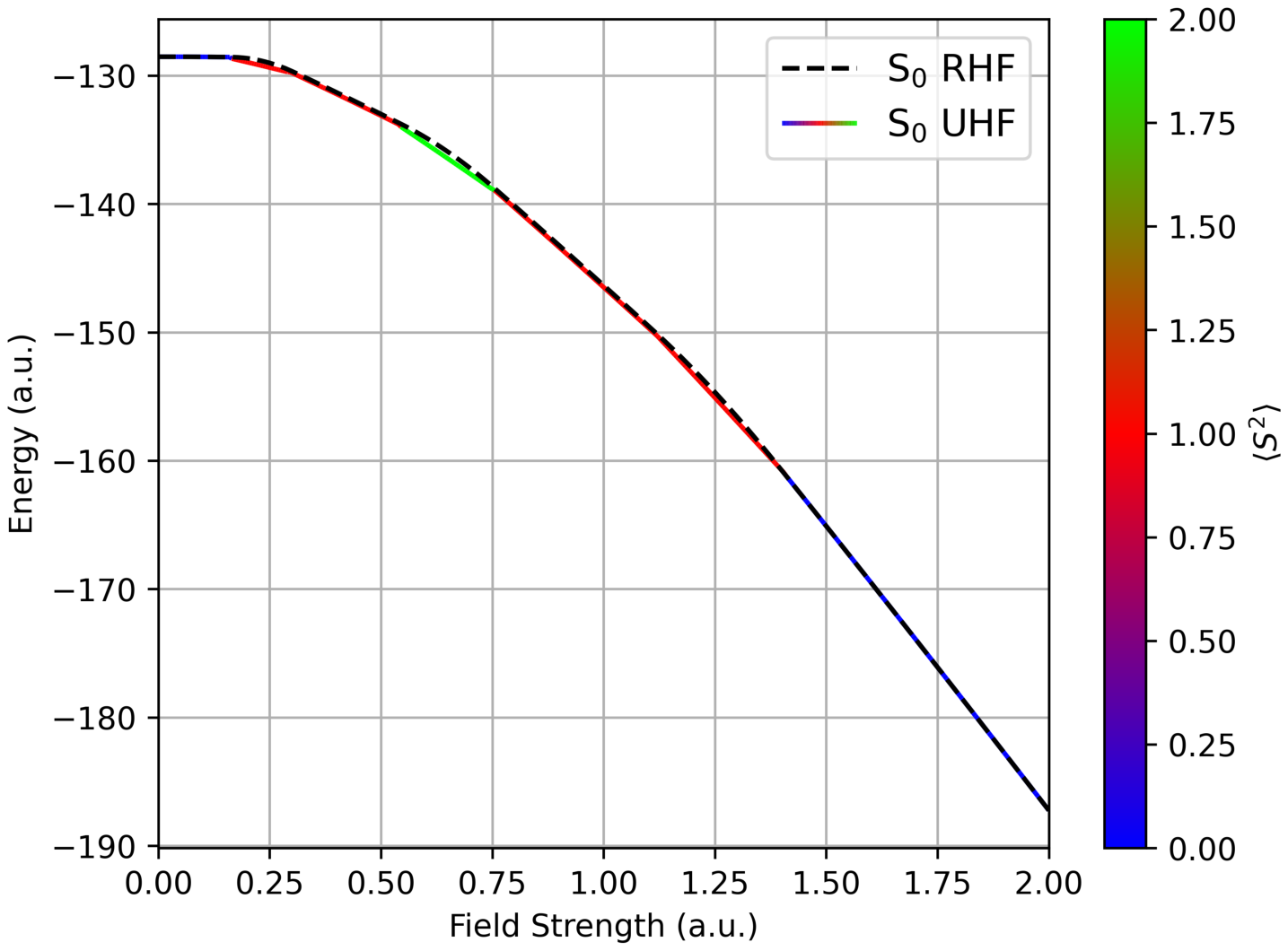}
    }
    \hfill
    \subfloat[Dipole moment \label{fig:ne_hf_dipole}]{%
         \includegraphics[scale=0.52]{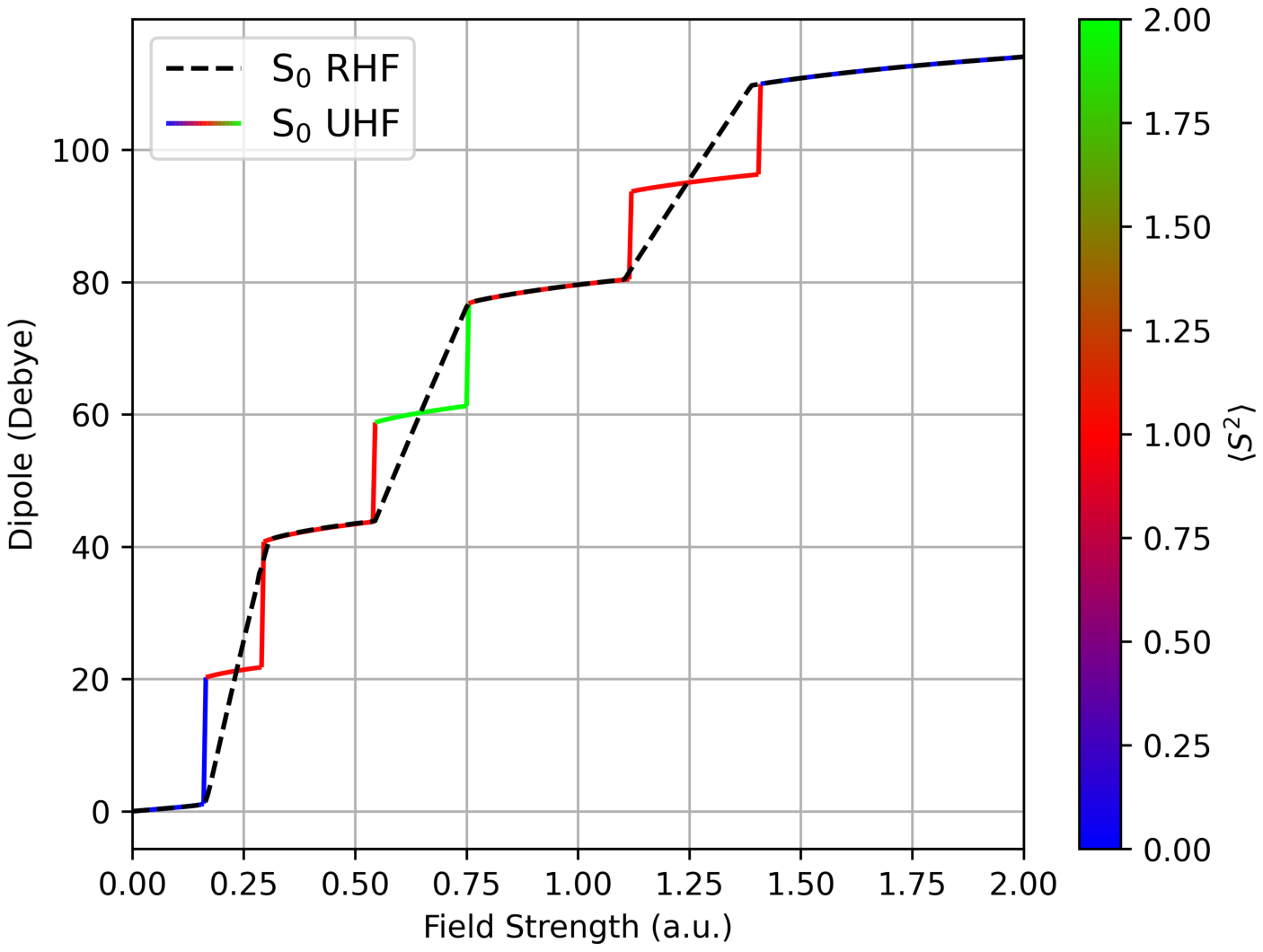}
    }

    \subfloat[Mulliken charges on Ne \label{fig:ne_hf_charge}]{%
         \includegraphics[scale=0.52]{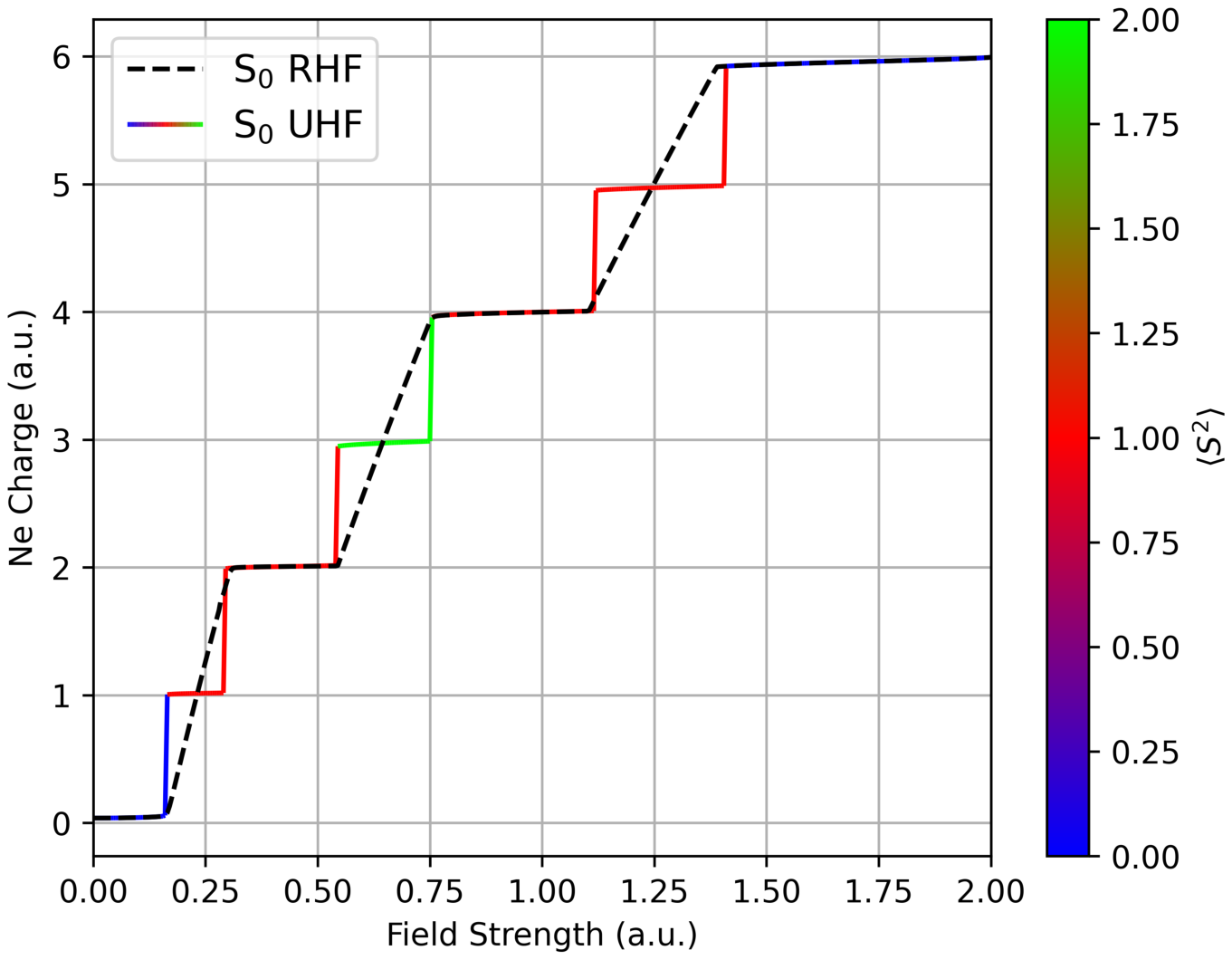}
    }
    \subfloat[$\Delta E^{atom}_{elec}$ \label{fig:ne_deip}]{%
         \includegraphics[scale=0.52]{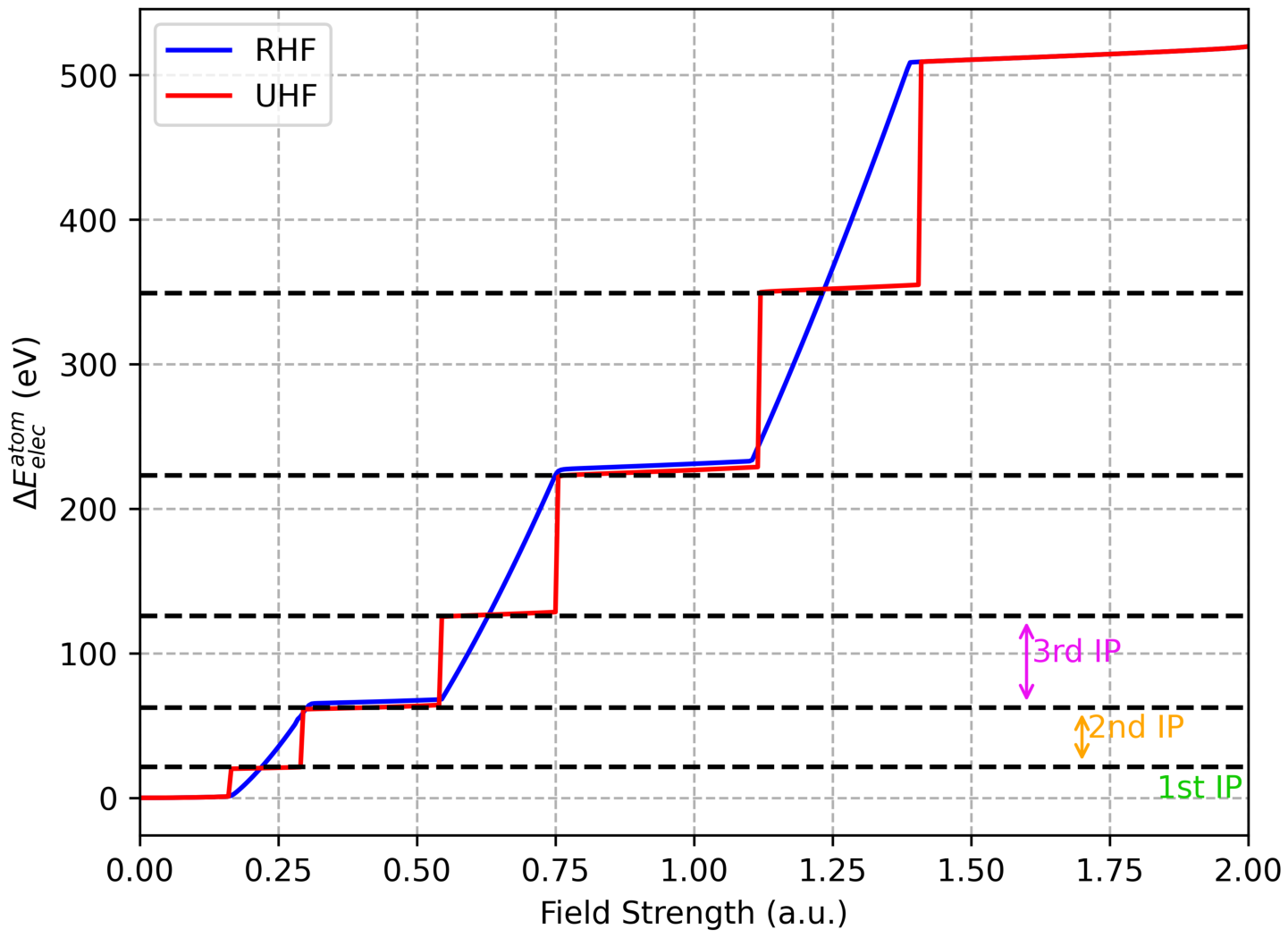}
    }
    \caption{ Hartree-Fock calculations of the field ionization of Ne using the aug-cc-pVTZ basis set augmented by hydrogenic 1s STO-3G functions at the ghost centers. (a) SCF total energy, (b) Dipole moment, (c) Mulliken charges on Ne and (d) $\Delta E^{atom}_{elec}$ as defined in Eq~\ref{eqn:deltaE_ip} as a function of field strength for the RHF and UHF solutions. In all cases, the RHF curve smoothly changes as the field strength increases, while UHF presents the characteristic kinks that are associated with ionization events. \lacrev{The IP's were calculated at the HF/aug-cc-pVTZ+ghost track level of theory.}}
    \label{fig:ne_all}
\end{figure}

In the previous section we thoroughly discussed how spin polarization is essential to qualitatively describe ionization by static fields in our model in which the continuum was discretized by a track of ghost functions placed along the direction of the applied electric field. At least for the He atom, we concluded that UHF properly captures, at the expense of spin contamination, the features associated with ionization within this limited model: we observe kinks in the potential energy curves as a function of field strength that are accompanied by discontinuities in the total electric dipole moment and the charge on the He atom. Now we move on to investigate if this simplest flavor of spin polarization (\emph{i.e.}, allowing $\alpha$ and $\beta$ orbitals to have different spatial parts) is enough to capture the features of atomic ionization in a heavier atom. Fig.~\ref{fig:ne_hf_energy} shows the RHF and UHF energy for the Ne atom. Once again, we see characteristic spin polarization and kinks in the UHF potential energy surface that are associated with ionization. We performed the same previous analysis for the electric dipole moment (Fig.~\ref{fig:ne_hf_dipole}) and Mulliken charges on the Ne atom (Fig.~\ref{fig:ne_hf_charge}) and they both agree with our previous discussion: at the expense of losing information about the total spin of the system (\emph{i.e.} spin contamination and our wavefunction not being an eigenstate of the $\hat{S}^2$ operator), we can now ionize an odd number of electrons for Ne through UHF, whereas those states are not accessible through a RHF reference.

The electronic structure of Ne is more complicated than He which also allows us to analyze other qualitative features of the ionized states of the system as the strength of the applied field increases. With this is mind, it is important to point out what one would expect for each of the Ne ionized states within this limited discretized continuum model that we used in our analysis. Table~\ref{tab:ne_ion_states} summarizes simplified electronic configurations expected for Ne, Ne$^+$, Ne$^{2+}$ and Ne$^{3+}$ and the track of ghost functions (with net charge $0, -1, -2, -3$) that represent the continuum-like states. Even though a proper analysis of the nature of the real ionized states of the atom is hindered due to the single determinant nature of \lacrev{HF} theory, one can notice that the S$_1$ and S$_2$ states in Table~\ref{tab:ne_ion_states} corresponds to one of the components of the physical (field-free) $^2P$ state of the Ne cation and $^3P$ state of the Ne dication, respectively. Our goal now is to see how we can obtain qualitative information about these physical states through Hartree-Fock theory within our limited model for ionization. 

\begin{table}[!htb]
    \centering
    \begin{tabular}{ |c | c | c|} \hline
    Ionization State & Ne configuration (only $2p$ orbitals) & Ghost configuration \\ \hline
    $\ket{S_0}$ (neutral) & $\ket{2p_x^2 2p_y^2 2p_z^2}$ & 
    $\ket{0}$ \\
    $\ket{S_1}$ (singly) & $\ket{2p_x^2 2p_y^2 2p_z^\alpha}$ & 
    $\ket{C_1^\alpha}$ \\     
    $\ket{S_2}$ (doubly) & $\ket{2p_x^2 2p_y^\alpha 2p_z^\alpha}$ & 
    $\ket{C_1^2}$ \\
    $\ket{S_3}$ (triply) & $\ket{2p_x^\alpha 2p_y^\alpha 2p_z^\alpha}$ & 
    $\ket{C_1^2 C_2^\alpha}$ \\\hline     
    \end{tabular}
    \caption{Simplified electronic configuration of Ne and ghost basis functions for the neutral atom and the first 3 ionized states $\ket{\textrm{S}_i}$. \lacrev{Each C$_i$ represents a ghost orbital that can be populated by ionized electrons.}}
    \label{tab:ne_ion_states}
\end{table}

\begin{figure}[!htb]
    \centering
    \subfloat[NOONs for Ne + ghost complex system \label{fig:ne_f250_no}]{%
         \includegraphics[scale=0.52]{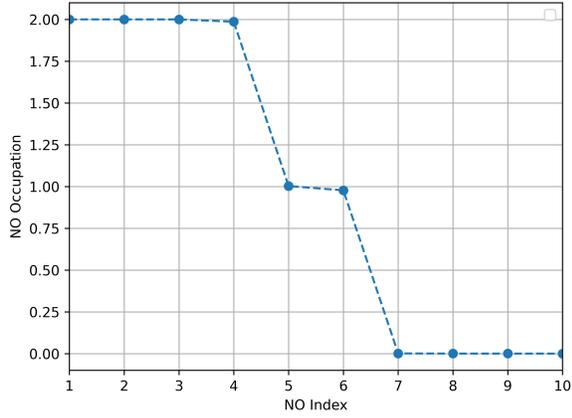}
    }
    \hfill
    \subfloat[NOONs for Ne subsystem \label{fig:ne_f250_nos}]{%
         \includegraphics[scale=0.52]{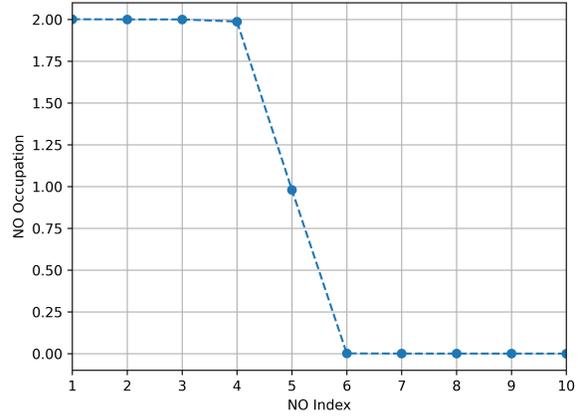}
    }

    \subfloat[NOONs for ghost subsystem \label{fig:ne_f250_nog}]{%
         \includegraphics[scale=0.52]{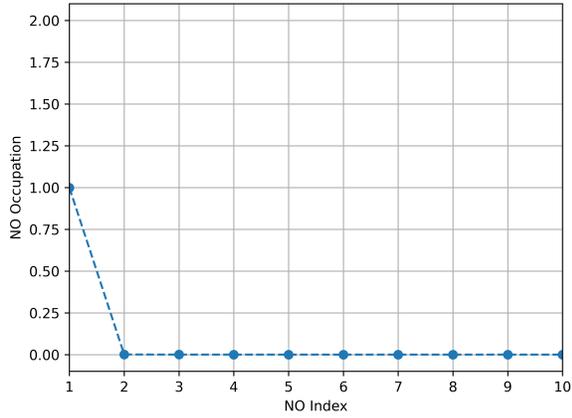}
    }
    \hfill
    \subfloat[UHF spin density \label{fig:ne_f250_spind}]{%
         \includegraphics[scale=0.52]{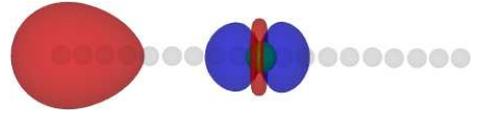}
    }
    \caption{Analysis of UHF for the singly ionized regime of Ne (F = 0.25 a.u.), using NOONs for (a) Ne + Ghost complex system, (b) Ne subsystem (c) Ghost subsystem; (d) UHF spin density plotted for an isovalue of 0.002 a.u. (blue and red parts indicate an excess of $\alpha$ and $\beta$ electrons respectively).}
    \label{fig:ne_f250_all}
\end{figure}

For the first ionization, we expect one unpaired electron at both the atom and the ghost functions, leading to a spin-contaminated UHF state in a similar fashion to what was described for the He case (Fig.~\ref{fig:he_no_all}). Fig.~\ref{fig:ne_f250_all} summarizes some of the features of the UHF for a field strength of $F = 0.25$ a.u., which lies within the region for single ionization. We can identify two unpaired electrons for the whole system (Fig.~\ref{fig:ne_f250_no}). Performing an eigenvalue analysis on the partitioned density matrices for the atom subsystem ($D^A$) and the continuum/ghost subsystem($D^G$) to obtain the natural orbital occupation numbers (NOONs) reveals that, as expected, each subsystem has a single unpaired electron (Figs.~\ref{fig:ne_f250_nos} and \ref{fig:ne_f250_nog}): an $\alpha$ electron for the discretized continuum and an excess $\beta$ electron for the atom (Fig.~\ref{fig:ne_f250_spind}), keeping $M_s = 0$ for the contaminated ground state.

\begin{figure}[!htb]
    \centering
    \subfloat[NOONs for Ne + ghost complex system \label{fig:ne_f500_no}]{%
         \includegraphics[scale=0.52]{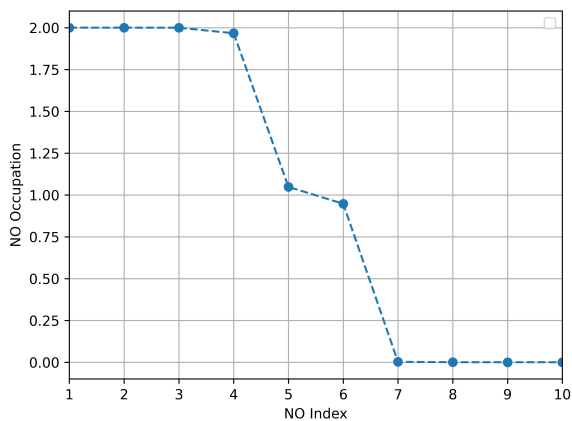}
    }
    \hfill
    \subfloat[NOONs for Ne subsystem \label{fig:ne_f500_nos}]{%
         \includegraphics[scale=0.52]{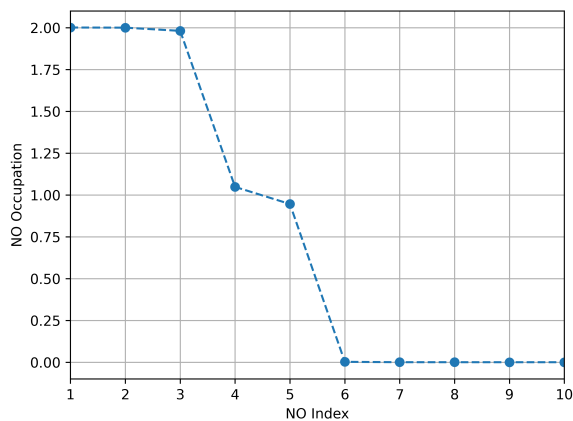}
    }

    \subfloat[NOONs for ghost subsystem \label{fig:ne_f500_nog}]{%
         \includegraphics[scale=0.52]{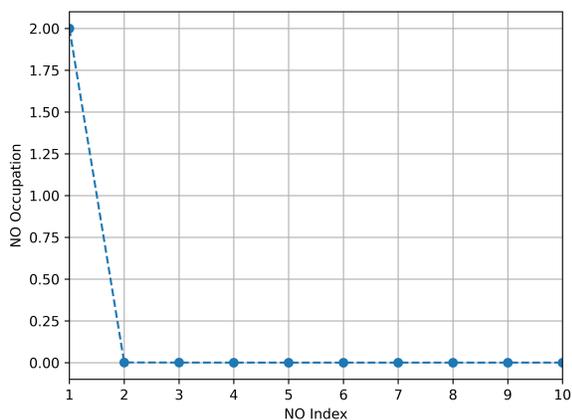}
    }
    \hfill
    \subfloat[UHF spin density,\label{fig:ne_f500_spind}]{%
         \includegraphics[scale=0.52]{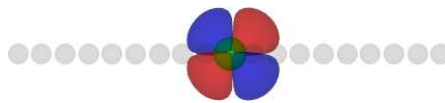}
    }
    \caption{Analysis of UHF for the doubly ionized regime of Ne (F = 0.5 a.u.), using  Natural Orbital Occupation Numbers (NOONs) for (a) Ne + Ghost complex system, (b) Ne subsystem (c) Ghost subsystem; (d) UHF spin density plotted for an isovalue of 0.002 a.u. (blue and red parts indicate an excess of $\alpha$ and $\beta$ electrons respectively).}
    \label{fig:ne_f500_all}
\end{figure}

The doubly ionized state in He simply corresponds to the pairing of the two electrons in one of the ghost functions, leaving behind just the atomic nucleus (Fig.~\ref{fig:he_no_all}, the curve indicated by a field strength of F = 0.5 a.u.). For Ne, however, we expect the second ionized electron to pair with the first one and localize around a ghost center while Ne becomes a triplet state characterized by two unpaired electrons with the same spin and consequently total $M_s = \pm 1$ (see Table~\ref{tab:ne_ion_states}). This state, however, cannot be achieved by following the UHF potential energy curve which is constrained to the $M_s = 0$ subspace. Instead, the UHF solution for this doubly ionized state of Ne is characterized by a $\langle S^2 \rangle= 1$ spin contaminated state comprised of an $\alpha$ electron and a $\beta$ unpaired electrons on the atom and no unpaired electrons for the track of ghost functions, as illustrated in Fig.~\ref{fig:ne_f500_all}. Some physical insight can be gained here by analyzing which states of the field-free Ne dication are included in the UHF $M_s$ = 0 field dependent solution in the doubly ionized regime: the external field, breaking spherical symmetry, mixes the $^1S$ and $^1D$ states of Ne$^{2+}$. Moreover, due to spin symmetry-breaking in UHF, we observe that one of the components ($M_s = 0$) of the $^3P$ is also included in the mixture. It should be noticed that such behavior also highlights one of the limitations of our approach to describe the ionized flux: by constraining electron density to escape the atom through a unique track of ghost functions, we observe a driving force towards pairing of the ionized electrons. This prevents neon from achieving the desired and expected triplet configuration. A possible, yet not explored in this work, way to minimize this issue would be to expand our description of the discretized continuum by adding more ghost functions into two separate parallel tracks \lacrev{(as illustrated by the grey crosses in Fig.~\ref{fig:model_setup})}, which would possibly allow the delocalization of the ionized electrons, leaving the atom subsystem with $M_s = \pm 1$ and the ghost subsystem with $M_s \mp 1$, while maintaining the total $M_s = 0$ for the UHF solution. 

Alternatively, since the main interest should be to capture as much physical insight for the atomic subsystem as possible within our limited model, it is possible to remove the $M_s = 0$ constraint in UHF by exploring the manifold of generalized Hartree-Fock solutions (GHF). In GHF, all of the exact spin symmetries of the system  (namely $\hat{S}^2$ and $\hat{S}_z$) can be broken by taking spin-orbitals that no longer belong to two independent sets of $\alpha$ and $\beta$ orbitals, but now are linear combinations of orbitals in both sets. This allows \lacrev{the spin-quantization axis of individual electrons} to rotate and not necessarily be parallel \lacrev{to each other} in order to achieve the variationally lowest possible energy state. This extra flexibility does not mean that the GHF solution cannot be obtained through UHF though. A GHF wavefunction is said to be non-collinear (and sometimes referred to as a true GHF solution) if its spin cannot be quantized along some axis and this illustrates a real UHF $\rightarrow$ GHF instability of the orbital hessian.\cite{mhg2015ghf,goings2015stability} This is the case for the double bond dissociation in $\textrm{CO}_2$ \cite{scuseria2011co2,mhg2015ghf} and for systems presenting spin-frustration.\cite{stetina2019modeling,yamaki2001generalized}. If such a spin quantization axis exists, the GHF solution can be found by exploring UHF wavefunctions with different spin multiplicity or lying within different $M_s$ manifolds. One can then find the difference between the number of up and down electrons by projecting the spin operators in this quantized axis.\cite{mhg2015ghf} In this context, we see that GHF can provide a convenient way to adiabatically follow the lowest UHF solution independently of $M_s$ constraints.

\begin{figure}[!htb]
    \centering
    \subfloat[Potential energy surfaces \label{fig:ne_ghf_pes}]{%
         \includegraphics[scale=0.52]{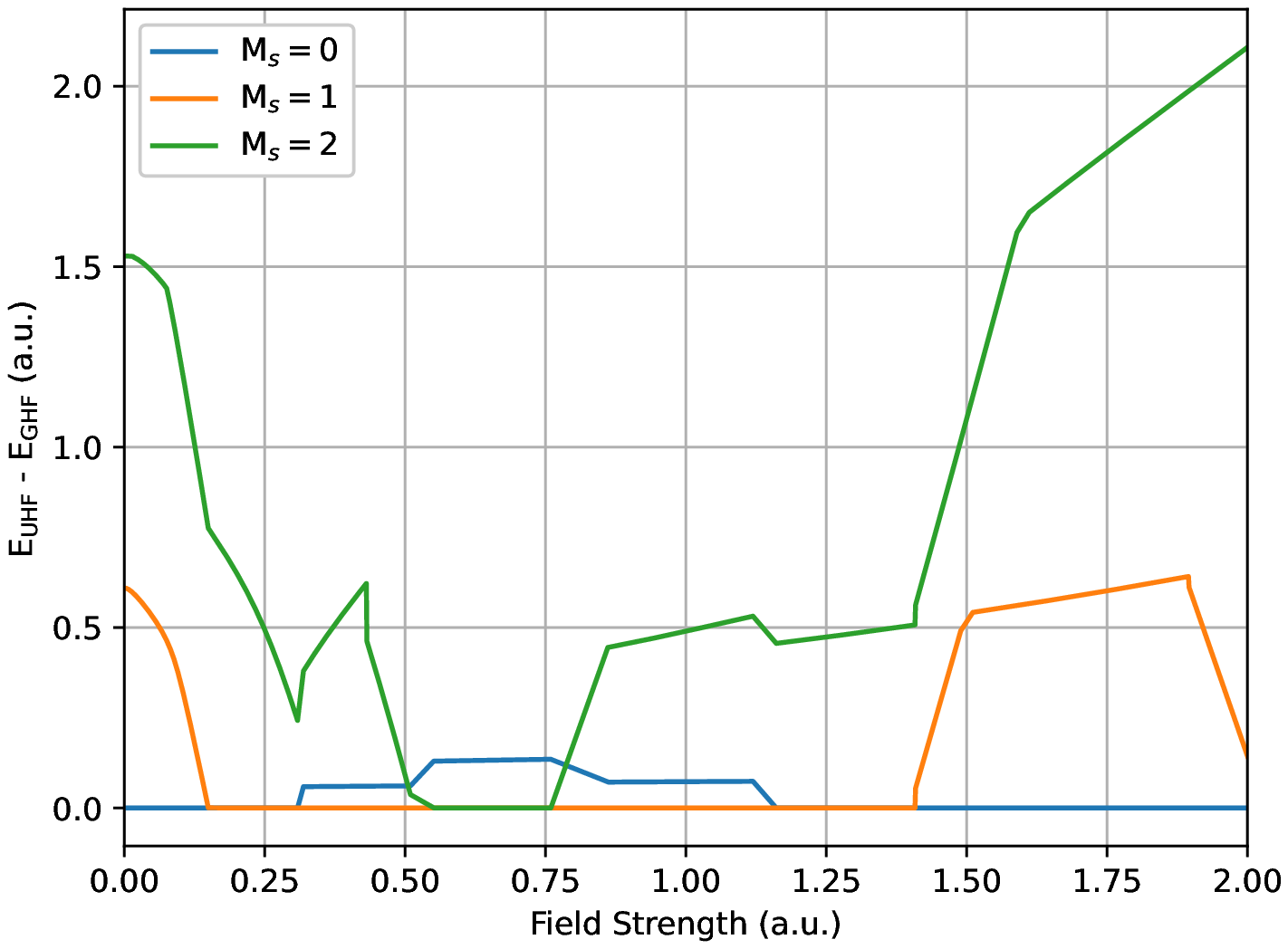}
    }
    \hfill
    \subfloat[Supersystem GHF effective number of up and down electrons \label{fig:ne_ghf_eps}]{%
         \includegraphics[scale=0.52]{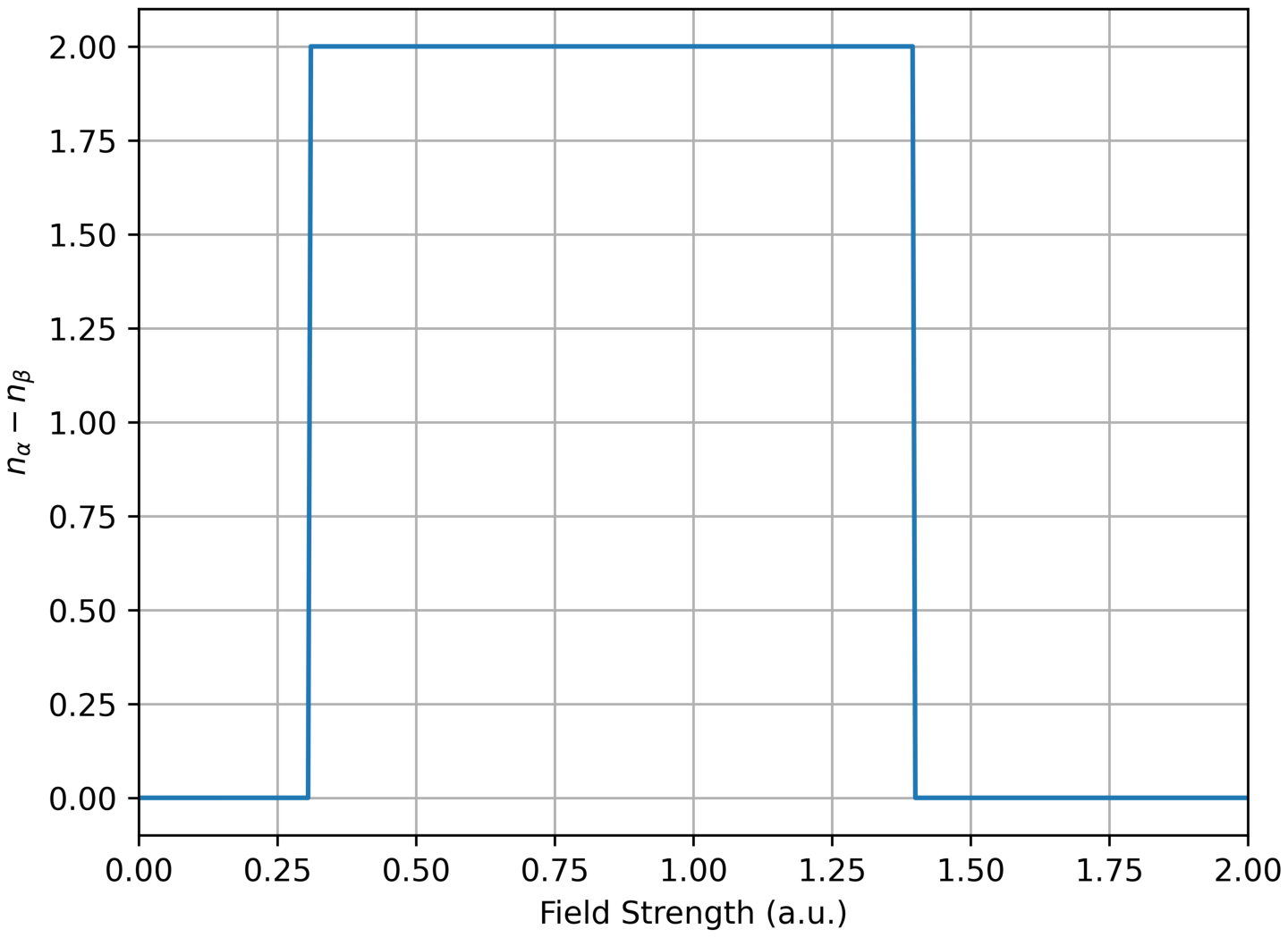}
    }
    \caption{(a) Potential energy curves as a function of field strength for Ne. \lacrev{It is possible to notice that, depending of the field strength, different M$_s$ UHF solutions become the lowest in energy and GHF connects these different subspaces by variationally targeting the unconstrained lowest energy state. (b) Difference between number of up and down electrons along the spin quantization axis for the GHF solution, indicating that for the doubly ionized regime (between F = 0.309 a.u. and F = 0.60 a.u), we have two unpaired $\alpha$ electrons left on the Ne dication.}}
    \label{fig:ghf_neon}
\end{figure}

Figure~\ref{fig:ne_ghf_pes} illustrates the behavior of the \lacrev{different M$_s$ UHF solutions compared to} GHF for the strong field ionization of the neon atom within the race track of ghost functions to represent discretized continuum states that we have been using so far. Even though we cannot notice significant differences between UHF and GHF in terms of their description of charges, dipoles and total number of unpaired electrons for the supersystem, the GHF potential energy curve follows the lowest UHF solution, \lacrev{highlighting its collinear nature}. \lacrev{For instance, for field strengths lower than F = 0.15 a.u., we observe that the M$_s$ = 0 UHF solution is the one with the lowest energy and that there is no difference between this solution and GHF, whereas the other M$_s$ states are higher in energy. For electric fields ranging from F = 0.15 a.u. and F = 0.309 a.u. (single ionization), however, the M$_s$ = 0 and M$_s$ = 1 states become degenerate, characterizing the RHF$\rightarrow$UHF instability previously discussed. For the regime between F = 0.309 a.u. and F = 0.60 a.u (which corresponds to the doubly ionized regime) the M$_s$ = 1 UHF solution becomes the lowest in energy, illustrating the transition between different M$_s$ subspaces that is captured naturally through GHF.} Our first UHF analysis was constrained within the $M_s = 0$ subspace and, therefore, the wrong qualitative behavior (one $\alpha$ and one $\beta$ electron) was observed for the double ionized state of the atom. GHF however, abolishes the $M_s = 0$ constraint and we can access the relevant doubly ionized state lying in the $M_s = \pm 1$ manifold (Fig.~\ref{fig:ne_ghf_eps}) and Ne achieves the expected configuration with two unpaired $\alpha$ electrons which better resembles the $M_s = \pm 1$ component of the physical $^3P$ field-free state of the Ne dication.

Finally, even though the plateau regions in Figs.~\ref{fig:ne_hf_dipole} and \ref{fig:ne_hf_charge} are well characterized by collinear GHF solutions that represent UHF wavefunctions with different $M_s$ values, the transition between these ionization states might not present the same behavior. At the onset of ionization, we suspect that, similarly to what was observed for the RHF $\rightarrow$ UHF instability, one would need a non-collinear GHF solution to properly describe the transition between one ionized state to the other. However, due to the dependence of the sharpness of this transition on the size of the artificial box delimited by the ghost functions, we did not pursue further investigation of this question. \lacrev{Hence, we note that GHF was used merely as a tool to access the lowest UHF state possible without any initial constraints on the number of $\alpha$ and $\beta$ electrons.}

\section{\label{sec:conclusions}Conclusions and Outlook}
In conclusion, the present work explored an analogy between the Coulson-Fischer point that has been extensively characterized for bond dissociation problems and static strong field ionization phenomena. Such extension is, at first, hindered by the fact that ionization implies an outgoing electron flux that is poorly described by the atom-centered Gaussian basis sets commonly used in electronic structure packages.\cite{QCHEM4} Hence, our basic ``race-track" model was introduced to account for this ionized flux by adding ghost basis functions spanning an appropriate region of space to allow some characterization of the continuum-like states of the ionized system. 

A preliminary analytical analysis of the ionization problem of He in a minimum basis with a single distant ghost function hinted that ionization events, which would be characterized by the localization of the electron density at one of the auxiliary ghost centers of the ``race-track", would not be well characterized by a restricted Hartree-Fock approach. In particular, the RHF solution could describe the neutral and the doubly ionized states of the system, but would lack any information about the singly ionized state given its electron-pairing constraint. This analysis also indicated the existence of a RHF $\rightarrow$ UHF instability for the singly ionized range of field strengths that was akin to the triplet instability that leads to spin polarization in the bond dissociation problem. The predictions from this toy model were validated by computations in a larger basis set. Moreover, the UHF potential energy surface as a function of field strength reveals kinks that, after an analysis of the charge population of the atom, the dipole moment of the system and the expectation value of the $\hat{S}^2$ operator, indicate the existence of an analog of the Coulson-Fischer point for static strong field ionization. We have also made an effort to eliminate possible spurious contributions stemming from our limited model of the discretized space by devising a partitioning scheme based on taking traces of the RHF and UHF density matrices over the appropriate basis functions centered only around the atom. Such scheme allowed us to, once again, determine the character of each state of the atom for different field strengths, as well as to recover information about the ionization energies of He. While RHF cannot represent the singly ionized state, UHF recovers the first and second ionization energies of He at the cost of breaking spin symmetry and being spin contaminated.

We then extended our analysis to a more complicated system: the Ne atom. The main takeaways are still valid: by constraining the HF solution to be an eigenfunction of $\hat{S^2}$, \emph{i.e.} constraining ourselves to the RHF manifold of solutions, we cannot describe ionization of an odd number of electrons as the strength of the external field increases. By exploring the solutions contained in the spin-polarized UHF manifold, on the other hand, we have a qualitatively correct description of the atomic charge associated with these ionized states. However, due to a strong drive to pair the electrons at the end of the auxiliary ``race-track" of ghost functions and the constraint imposed on the value of $\langle \hat{S}_z \rangle$ for UHF solutions, the results do not correspond to the expected Hund's rule configuration of Ne for some field strengths. Thus, we also analyzed the use of GHF solutions to obtain a better qualitative description of the static strong field ionization of Ne. We observed that, while maintaining the same behavior as UHF for the atomic charges, GHF can properly rotate spins at the cost of losing information about the expectation value of $\hat{S}_z$ in order to give the proper high-spin description of the unpaired electrons on the atomic subsystem. Exploring how different kinds of symmetry-broken solutions, such as allowing complex polarization in Hartree-Fock (cRHF, cUHF and cGHF),\cite{ostlund1972complex,small2015restricted,lee2019two,lee2019cr} could also prove potentially useful to obtain qualitatively correct description of strong field ionization of more complex systems. 

Even though the results and analysis presented in the current work were obtained for the case of an applied external static electric field, it can be argued that some of the consequences of imposing spin constraints on the Hartree-Fock solutions could also lead to a qualitatively wrong description of the real-time dynamics of the system when a physically meaningful laser pulse is used. For the more common problem of bond stretching in electronic structure, it has already been shown that using a qualitatively bad HF solution as the starting point for more elaborate methods that account for dynamic electron correlation could lead to failures in obtaining a good description of potential energy surface, for example. In this sense, work towards extending the analysis here to real-time time-dependent electronic structure is underway in our group. To achieve more physically meaningful results, a better approach to represent the continuum states, perhaps through a mixed gaussian-plane wave basis set or complex absorbing potentials, is needed.\cite{pw2011,nimrod2020}

\section*{Supplementary Material}
See supplementary material for the raw data (energies, dipole, charges) organized in XLXS format and additional plots (PDF).

\begin{acknowledgments}
This research was supported by the Director,  Office of Science,  Office of Basic Energy Sciences, of the U.S. Department of Energy under Contract No. DE-AC02-05CH11231. L.A.C. would like to thank Dr. Josh Cantin and Juan Arias-Martinez for stimulating discussions. M.H.-G. is a part-owner of Q-Chem, which is the software platform in which the computations described here were carried out.
\end{acknowledgments}

\section*{Data Availability}
The data that supports the findings of this study are available within the article and its supplementary material.

\bibliography{refs}

\end{document}